\documentclass[12pt]{article}
\usepackage{amsfonts}
\usepackage{amsmath}
\usepackage{amsthm}
\usepackage{graphicx}
\usepackage{epstopdf}
\usepackage{caption}
\usepackage{subcaption}
\usepackage[right=2.1cm,left=2.1cm, top=2.5cm, bottom=2.5cm]{geometry}

\title{On Bayesian inference for the M/G/1 queue \\ with efficient MCMC sampling}
\author{Alexander Y. Shestopaloff \\
Department of Statistical Sciences \\
University of Toronto \\
alexander@utstat.utoronto.ca \\
\and Radford M. Neal \\
Department of Statistical Sciences \\
\& Department of Computer Science \\
University of Toronto \\
radford@utstat.utoronto.ca}
\date{31 December 2013}

\begin{document}

\maketitle

\begin{abstract}
We introduce an efficient MCMC sampling scheme to perform Bayesian inference in the M/G/1 queueing model given only observations of interdeparture times. Our MCMC scheme uses a combination of Gibbs sampling and simple Metropolis updates together with three novel ``shift'' and ``scale'' updates. We show that our novel updates improve the speed of sampling considerably, by factors of about $60$ to about $180$ on a variety of simulated data sets. 
\end{abstract}

This paper proposes a new approach to computation for Bayesian inference for the M/G/1 queue (Markovian arrival process/General service time distribution/1 server). Inference for this model using ABC (Approximate Bayesian Computation) was previously considered by Bonassi (2013), Fearnhead and Prangle (2012), and Blum and Francois (2010). ABC, in general, does not yield samples from the exact posterior distribution. We use the strategy of considering certain unobserved quantities as latent variables, allowing us to use Markov Chain Monte Carlo (MCMC), which converges to the exact posterior distribution. 

\section{The model}

In the M/G/1 queueing model, customers arrive at a single server with independent interarrival times, $W_{i}$, distributed according to the $\textnormal{Exp}(\theta_{3})$ distribution. Here, $\theta_{3}$ is the arrival rate, hence $W_{i}$ has density function $f(w_{i}) = \theta_{3} \exp(-\theta_{3}w_{i})$ for $w_{i} \geq 0$ and $0$ otherwise. They are served with independent service times $U_{i}$, which have a $\textnormal{Uniform}\left(\theta_{1}, \theta_{2}\right)$ distribution. (Our
 MCMC approach can be generalized to other service time distributions.) We do not observe the interarrival times, only the interdeparture times, $Y_{i}$. The goal is to infer the unknown parameters of the queueing model, $\theta = (\theta_{1}, \theta_{2}, \theta_{3})$, using observed interdeparture times, $y = (y_{1}, \ldots, y_{n})$. We assume that the queue is empty before the first arrival.

The process of interdeparture times can be written as
\begin{eqnarray}
Y_{i} &=& U_{i} + \max\biggl(0, \sum_{j = 1}^{i}W_{j} -  \sum_{j = 1}^{i-1} Y_{j}\biggr) \ = \ U_{i} + \max(0, V_{i} -  X_{i-1})
\end{eqnarray}
where $W_{j}$ is the time from the arrival of customer $j-1$ (or from 0 when $j = 1$) to the arrival of customer $j$, $V_{i} = \sum_{j = 1}^{i}W_{j}$ is the arrival time of the $i$-th customer, and $X_{i} = \sum_{j = 1}^{i} Y_{i}$ is the departure time of the $i$-th customer, with $X_{0}$ defined to be $0$.

We take the arrival times, $V_{i}$, to be latent variables. The $V_{i}$ evolve in time as a Markov process. Viewed this way, the queueing model can be summarized as follows:
\begin{eqnarray}
V_{1} &\sim& \textnormal{Exp}(\theta_{3}) \\
V_{i} | V_{i-1} &\sim& V_{i-1} +  \textnormal{Exp}(\theta_{3}),  \quad  i = 2, \ldots, n \\
Y_{i} | X_{i-1}, V_{i} &\sim& \textnormal{Uniform}(\theta_{1} + \max(0, V_{i} - X_{i - 1}), \ \theta_{2} + \max(0, V_{i} - X_{i - 1})), \quad i = 1, \ldots, n
\end{eqnarray}
For convenience, set $v = (v_{1}, \ldots, v_{n})$. The joint density of the $V_{i}$ and the $Y_{i}$ can be factorized as follows
\begin{eqnarray}
P(v, y | \theta) = P(v_{1} | \theta)\prod_{i = 2}^{n}P(v_{i}|v_{i - 1}, \theta)\prod_{i = 1}^{n}P(y_{i} | v_{i}, x_{i-1}, \theta)
\end{eqnarray}
Let $\pi(\theta)$ be a prior for $\theta$. The posterior distribution $\pi(\theta | y)$ of $\theta$ is then
\begin{eqnarray}
\pi(\theta|y) &\propto& \pi(\theta)\int \cdots \int  P(v_{1} | \theta)\prod_{i = 2}^{n}P(v_{i}|v_{i - 1}, \theta)\prod_{i = 1}^{n}P(y_{i} | v_{i}, x_{i-1}, \theta) dv_{1} \cdots dv_{n}
\label{eq:vxpdf}
\end{eqnarray}
The integral (\ref{eq:vxpdf}) cannot be computed analytically. Hence to sample from $\pi(\theta | y)$ we need to include the $V_{i}$ in the MCMC state and sample from the joint posterior distribution of the $V_{i}$ and $\theta$ given the data, which is
\begin{eqnarray}
\pi(v, \theta | y) &\propto& \pi(\theta)P(v_{1} | \theta)\prod_{i = 2}^{n}P(v_{i}|v_{i - 1}, \theta)\prod_{i = 1}^{n}P(y_{i} | v_{i}, x_{i-1}, \theta)
\label{eq:vxpost}
\end{eqnarray}
Taking the values of $\theta$ from each draw from (\ref{eq:vxpost}) and ignoring the $V_{i}$ will yield a sample from the posterior distribution of $\theta$.

\section{MCMC sampling}

Our MCMC procedure will be based on combining five different updates. The first is a Gibbs update that draws new values for each $V_{i}$, given values of the other $V_{i}$, $\theta$, and the data. The second is a simple Metropolis update for $\theta$, given values of the $V_{i}$ and the data. The last three updates are novel --- one Metropolis ``shift'' update and two Metropolis-Hastings-Green ``scale'' updates that propose to simultaneously change components of $\theta$ and all of the $V_{i}$. 

\subsection{Gibbs updates for arrival times}

We first work out how to apply standard Gibbs sampling updates for the $V_{i}$, for which we need to derive the full conditional density for each $V_{i}$ given the other $V_{i}$ and $\theta$. We denote all of the $V_{j}$ except $V_{i}$ as $V_{-i}$. We consider three cases, when $i = 1$, when $2 \leq i \leq n - 1$ and when $i = n$.

For the case $i = 1$, we have
\begin{eqnarray}
P(v_{1} | v_{-1}, y, \theta) &\propto& P(v_{1})P(y_{1} | v_{1}, \theta)P(v_{2} | v_{1}, \theta)\nonumber \\
&\propto& \theta_{3}e^{-\theta_{3} v_{1}} I(0 \leq v_{1}) \frac{I(y_{1} \in [\theta_{1} + v_{1}, \theta_{2} + v_{1}])}{\theta_{2} - \theta_{1}}\nonumber \\ 
&& \times \ \theta_{3}e^{-\theta_{3} (v_{2} - v_{1})} I(v_{1} \leq v_{2}) \nonumber \\
&\propto& I(v_{1} \in [\max(0, x_{1} - \theta_{2}), \min(v_{2}, x_{1} - \theta_{1})])
\end{eqnarray}
So, conditional on the parameters and the observed data, the distribution of $V_{1}$ is $\textnormal{Uniform}(\max(0, x_{1} - \theta_{2}), \min(v_{2}, x_{1} - \theta_{1}))$.

When $2 \leq i \leq n - 1$, we have
\begin{eqnarray}
P(v_{i} | v_{-i}, y, \theta) &\propto& P(v_{i} | v_{i-1}, \theta)P(y_{i} | x_{i-1}, v_{i}, \theta)P(v_{i+1} | v_{i}, \theta) \nonumber\\
&\propto&\theta_{3}e^{-\theta_{3} (v_{i} - v_{i - 1})}I(v_{i-1} \leq v_{i})\notag \\
&& \times \ \frac{I(y_{i} \in [\theta_{1} + \max(0, v_{i} - x_{i - 1}), \theta_{2} + \max(0, v_{i} - x_{i - 1})])}{\theta_{2} - \theta_{1}}\nonumber \\
&& \times \ \theta_{3}e^{-\theta_{3} (v_{i + 1} - v_{i})}I(v_{i} \leq v_{i+1}) \notag \\
&\propto& I(v_{i} \in [v_{i-1}, v_{i+1}])I(y_{i} \in [\theta_{1} + \max(0, v_{i} - x_{i - 1}), \theta_{2} + \max(0, v_{i} - x_{i - 1})]) \ \ \ \ \ \
\label{eq:innercase}
\end{eqnarray}
To simplify this expression, note that $x_{i} = y_{i} + x_{i-1}$, and first consider the case $y_{i} > \theta_{2}$. When this is so, we must have $v_{i} > x_{i - 1}$, hence, in this case $V_{i}$ will have a Uniform distribution on $[x_{i} - \theta_{2}, \min(v_{i + 1}, x_{i} - \theta_{1})]$. Now consider the case $y_{i} \leq \theta_{2}$. We rewrite expression (\ref{eq:innercase}) as
\begin{eqnarray}
\lefteqn{I(v_{i} \in [v_{i-1}, v_{i+1}])I(y_{i} \in [\theta_{1} + \max(0, v_{i} - x_{i - 1}), \theta_{2} + \max(0, v_{i} - x_{i - 1})])I(v_{i} \leq x_{i-1})} \nonumber \\
&+& I(v_{i} \in [v_{i-1}, v_{i+1}])I(y_{i} \in [\theta_{1} + \max(0, v_{i} - x_{i - 1}), \theta_{2} + \max(0, v_{i} - x_{i - 1})])I(v_{i} > x_{i-1}) \nonumber \\
&& = \ I(v_{i} \in [v_{i-1}, x_{i+1}]) + I(v_{i}  \in (x_{i-1}, \min(v_{i+1}, x_{i} - \theta_{1})]) \nonumber \\
&& = \ I(v_{i} \in [v_{i-1},  \min(v_{i+1}, x_{i} - \theta_{1})])
\end{eqnarray}
We see that for $y_{i} \leq \theta_{2}$, $V_{i}$ will have a Uniform distribution on $[v_{i - 1}, \min(v_{i + 1}, x_{i} - \theta_{1})]$.

Finally, for the case $i = n$, we have
\begin{eqnarray}
P(v_{n} | v_{-n}, y, \theta) &\propto& P(v_{n} | v_{n - 1})P(y_{n} | v_{n}) \nonumber \\
&\propto& e^{-\theta_{3} (v_{n} - v_{n - 1})}I(v_{n-1} \leq v_{n}) \nonumber \\
&&\times \ I(y_{n} \in [\theta_{1} + \max(0, v_{n} - x_{n - 1}),  \theta_{2} + \max(0, v_{n} - x_{n - 1})])
\end{eqnarray}
From this it follows that $V_{n}$ will have an Exponential distribution, truncated to $[x_{n} - \theta_{2}, x_{n} - \theta_{1}]$ when $y_{n} > \theta_{2}$ and to $[v_{n - 1}, x_{n} - \theta_{1}]$ when $y_{n} \leq \theta_{2}$.

All of the above distributions can be easily sampled from by using the inverse CDF method. The case $i = n$ deserves a small note. If we suppose the truncation interval of $V_{n}$ is $[L, U]$, then the CDF and inverse CDF will be
\begin{equation}
F_{V_{n}}(v_{n}) = \frac{e^{-\theta_{3}L} -  e^{-\theta_{3}v_{n}}}{e^{-\theta_{3}L} -  e^{-\theta_{3}U}}, \quad F_{V_{n}}^{-1}(y) = \log((1-y)e^{-\theta_{3}L} + ye^{-\theta_{3}U})
\end{equation}
So we can sample $V_{n}$ as $F_{V_{n}}^{-1}(R)$ where $R$ is Uniform$(0, 1)$.

\subsection{Simple Metropolis updates for parameters}

We use simple Metropolis updates (Metropolis, et al (1953)) to sample from the conditional posterior distribution of $\theta$ given values of the $V_{i}$ and the data, $\pi(\theta | v, y) \propto \pi(v, \theta | y)$. 

When doing simple Metropolis updates of $\theta$ given the arrival times and the data, we used the 1-to-1 reparametrization $\eta = (\eta_{1}, \eta_{2}, \eta_{3}) = (\theta_{1}, \theta_{2} - \theta_{1}, \log(\theta_{3}))$.

A simple Metropolis update for $\eta$ that leaves $\pi(\eta | v, y)$ invariant proceeds as follows. We choose a symmetric proposal density $q(\cdot | \eta)$, for which $q(\eta^{*} | \eta) = q(\eta | \eta^{*})$. Given the current value $\eta$, we generate a proposal $\eta^{*} \sim q(\eta^{*} | \eta)$. We compute the acceptance probability
\begin{eqnarray}
a = \min\biggl(1, \frac{\pi(\eta^{*} | v, y)}{\pi(\eta | v, y)}\biggr)
\label{eq:metacc}
\end{eqnarray}
and let the new state, $\eta'$, be $\eta^{*}$ with probability $a$ and otherwise reject the proposal, letting $\eta' = \eta$.

We use a normal proposal with independent coordinates centered at the current value of $\eta$, updating all components of $\eta$ at once. If the proposed value for $\eta_{1}$ or $\eta_{2}$ is outside its range, the proposal can be immediately rejected.

To prevent overflow or underflow, all MCMC computations use the logarithm of the posterior (\ref{eq:vxpost}), which (up to an additive constant) simplifies to
\begin{eqnarray}
\log \pi(v, \theta | y) &=&  \log(\pi(\theta)) + n\log(\theta_{3}) - \theta_{3} v_{n} - n\log(\theta_{2} - \theta_{1})
\label{eq:logpost}
\end{eqnarray}
whenever the following constraints are satisfied
\begin{eqnarray}
\label{eq:c1}
&& \theta_{2} - \theta_{1} > 0 \\
\label{eq:c2}
&& \theta_{1} \leq y_{1} - v_{1} \\
&& \theta_{2} \geq y_{1} - v_{1} \\
&& \theta_{1} \leq \min(y_{i} - \max(0, v_{i} - x_{i-1})) \quad \textnormal{for all} \ i \geq 2 \\
&& \theta_{2} \geq \max(y_{i} - \max(0, v_{i} - x_{i-1})) \quad \textnormal{for all} \ i \geq 2
\label{eq:c5}
\end{eqnarray}
Otherwise, $\log(\pi(v, \theta) | y)) = -\infty$.

The dominant operation when computing the log likelihood for a data set of length $n$ is checking the constraints (\ref{eq:c1}) to (\ref{eq:c5}), which requires time proportional to $n$. Storing the constraints on $\theta_{1}$ and $\theta_{2}$ given by (\ref{eq:c2}) to (\ref{eq:c5}) during evaluation of the log likelihood at $(\theta, v_{i})$ allows us to evaluate the log likelihood for another $\theta^{*}$ and the same $v_{i}$ in constant time. This allows us to do $K$ additional Metropolis updates for less than $K$ times the computational cost it takes to do a single update, which is likely to make sampling more efficient. A similar improvement in sampling should be possible for models with other service time distributions that have low-dimensional sufficient statistics. As well, doing additional simple Metropolis updates makes an imperfect choice of proposal distribution have less of an effect on sampling efficiency.

\subsection{Shift and scale updates}

The Gibbs and simple Metropolis updates are sufficient to give an ergodic MCMC scheme. However, as we will see, these updates are sometimes very inefficient when used on their own. In this section, we introduce our novel shift and scale updates, which can make MCMC sampling much more efficient.

Shift updates and scale updates are used to sample from the joint posterior distribution of the parameters and the latent variables, $\pi(v, \theta | y)$. The shift update takes the form of a standard Metropolis update, while the scale updates are Metropolis-Hastings-Green (MHG) updates. 

In general, an MHG update proceeds as follows. We introduce an extra variable $z \in \{-1, +1\}$. Given a current value $(v, \theta)$ and a density $q(\cdot | v, \theta)$ we generate $z \sim q(z | v, \theta)$. We then propose $(v^{*}, \theta^{*}, z^{*}) = g(v, \theta, z)$. The function $g$ is the inverse of itself, with a Jacobian $|\nabla_{(v, \theta)} g(v, \theta, z)|$ which is nowhere zero or infinite. We compute the acceptance probability
\begin{eqnarray}
a = \min\biggl(1, \frac{\pi(\theta^{*}, v^{*} | y)q(z^{*} | v^{*}, \theta^{*})}{\pi(\theta, v | y)q(z | v, \theta)}|\nabla_{(v, \theta)} g(v, \theta, z)| \biggr)
\label{eq:mhgacc}
\end{eqnarray}
and let the new state, $(v', \theta')$ be $(v^{*}, \theta^{*})$ with probability $a$ and let $(v', \theta') = (v, \theta)$ otherwise. For more on the MHG algorithm, see Geyer (2003).

The motivation for the shift updates and scale updates is that conditioning on given values of the arrival times constrains the range of parameter values for which the posterior density is non-zero, preventing us from changing the parameters by a significant amount. This can lead to inefficient sampling when $\theta$ is updated given the $V_{i}$. In contrast, our new shift and scale updates change the latent variables and the parameters simultaneously, in accordance with their dependence structure, thus allowing for much greater changes to the parameters.

Hard constraints on $\theta$ given $v$ aren't necessary for these updates to be beneficial.  If the service
time density were non-zero for all positive values, the distribution of $\theta$ given $v$ might still be much more concentrated than the marginal posterior for $\theta$.

\subsubsection{Shift updates}

We first consider updates that shift both the minimum service time, $\theta_{1}$, and all the arrival times, $V_{i}$, keeping the range of service times, $\theta_{2} - \theta_{1}$, fixed. Note that for $i = 1$ and for all $i$ when $X_{i-1} < V_{i}$, we have $\theta_{1} < X_{i} - V_{i}$. Hence the values of one or more of the $V_{i}$ will constrain how much we can propose to change $\theta_{1}$ --- any proposal to change $\theta_{1}$ that violates these constraints must be rejected.

A shift update addresses the presence of these constraints as follows. We first draw a shift $s$ from any distribution symmetric around $0$. Then, we propose new values $V_{i}^{*} = V_{i} - s$ for all $i$ and $\theta_{1}^{*} = \theta_{1} + s$. So, a proposal to increase the minimum service time is coupled with a simultaneous proposal to decrease all of the arrival times in a way that keeps the constraints between the arrival times and the minimum service time satisfied. 

A shift proposal will be rejected if it proposes a value for $V_{1}$ or $\theta_{1}$ that is less than $0$. Otherwise, the acceptance probability for a shift update is
\begin{eqnarray}
a =  \min\biggl(1, \frac{\pi(\theta^{*}, v^{*} | y)}{\pi(\theta, v | y)} \biggr)
\end{eqnarray}

\subsubsection{Range scale updates}

Next, we consider range scale updates that propose to simultaneously change the latent variables and the range of service times $\theta_{2} - \theta_{1}$, which is also constrained by the latent variables --- specifically, for $V_{1}$ and for all $i > 2$ where $V_{i} > X_{i-1}$, we have $\theta_{2} - \theta_{1} > X_{i} - \theta_{1} - V_{i}$. These updates propose to scale the ``gaps'' $X_{i} - \theta_{1} - V_{i}$ (gaps between the current arrival time and the latest possible arrival time), while at the same time proposing to scale $\theta_{2} - \theta_{1}$ by the same amount, keeping $\theta_{1}$ fixed.

In particular, we first fix (or draw from some distribution) a scale factor $c_{\textnormal{range}} > 0$. We then draw $z \sim \textnormal{Uniform}\{-1, 1\}$. The function $g(v, \theta, z)$ proposes to change $V_{i}$ to $V_{i}^{*} = (X_{1} - \theta_{1}) - c^{z}_{\textnormal{range}}(X_{i} - \theta_{1} - V_{i})$ and $\theta_{2} - \theta_{1}$ to $c^{z}_{\textnormal{range}}(\theta_{2} - \theta_{1})$. We set $z^{*} = -z$. The Jacobian $|\nabla_{(v, \theta)} g(v, \theta, z)| = c^{z(n+1)}_{\textnormal{range}}$.

A range scale proposal will be rejected if it proposes a set of $V_{i}$ for which some $V_{i} < V_{i-1}$ or for which $V_{1} < 0$. Otherwise, the MHG acceptance probability for a range scale update is
\begin{eqnarray}
a = \min\biggl(1, \frac{\pi(\theta^{*}, v^{*} | y)}{\pi(\theta, v | y)}c^{z(n+1)}_{\textnormal{range}} \biggr)
\end{eqnarray}

\subsubsection{Rate scale updates}

Finally, we consider rate scale updates that propose to simultaneously change both the interarrival times $W_{i} = V_{i} - V_{i-1}$, and the arrival rate $\theta_{3}$. We motivate these updates as follows. We expect the distribution of $\theta_{3}$ given the interarrival times $W_{i}$ to be concentrated around $1/\overline{W}$. Consequently, we would expect the joint posterior of $\theta_{3}$ and the $W_{i}$ to have a high density along a ridge with values $cW_{i}$ and $\theta_{3}/c$ for some $c$, as long as $cW_{i}$ and $\theta_{3}/c$ do not lie in region with low probability. So, proposing to change $W_{i}$ to $cW_{i}$ and $\theta_{3}$ to $\theta_{3}/c$ for some $c$ potentially keeps us in a region of high density.

We perform these updates in terms of $\eta_{3} = \log(\theta_{3})$. In detail, this update proceeds as follows. We first fix (or draw from some distribution) a scale $c_{\textnormal{rate}} > 0$. We then draw $z \sim \textnormal{Uniform}\{-1, 1\}$. The function $g(v, \eta, z)$ proposes to change all $W_{i}$ to $c^{z}_{\textnormal{rate}}W_{i}$ and $\eta_{3}$ to $\eta_{3} - \log(c^{z}_{\textnormal{rate}})$. We set $z^{*} = -z$. The Jacobian $|\nabla_{(v, \eta)} g(v, \eta, z)| = c^{zn}_{\textnormal{rate}}$. 

We reject the proposal immediately if it violates the following constraints: for $i$ where $y_{i} > \theta_{2}$ the proposed arrival time $V_{i}^{*} = \sum_{i=1}^{n}c^{z}_{\textnormal{rate}}W_{i}$ must be in $[x_{i} - \theta_{2}, x_{i} - \theta_{1}]$, and for $i$ where $y_{i} \leq \theta_{2}$, we must have $V_{i}^{*} \leq x_{i} - \theta_{1}$. Otherwise, the MHG acceptance probability for a rate scale update is
\begin{eqnarray}
a = \min\biggl(1, \frac{\pi(\eta^{*}, v^{*} | y)}{\pi(\eta, v | y)}c^{zn}_{\textnormal{rate}} \biggr)
\end{eqnarray}

\subsubsection{Ergodicity}

Although a scheme consisting of the shift, range scale, and rate scale updates changes all of the latent variables and parameters, it is not ergodic. To see this, consider updating the arrival times $V_{i}$, or equivalently, interarrival times $W_{i} = V_{i} - V_{i-1}$. Shift updates do not change the interarrival times. Range scale updates with scale factor $c_{\textnormal{range}}$ change each interarrival time as $W_{i}^{*} = y_{i}(1 - c^{z}_{\textnormal{range}}) + c^{z}_{\textnormal{range}}W_{i}$, and rate scale updates with scale factor $c_{\textnormal{rate}}$ change each interarrival time as $W_{i}^{*} = c^{z}_{\textnormal{rate}}W_{i}$. If we are in a situation where $W_{i} = W_{j}$ and $y_{i} = y_{j}$ for one or more $j \neq i$, then all subsequent updates will keep $W_{i} = W_{j}$. Note that this is true even if $c_{\textnormal{range}}$ and $c_{\textnormal{rate}}$ are randomly selected at each update.

Hence, our novel updates must still be combined with simple Gibbs sampling updates of the $V_{i}$'s to get an ergodic sampling scheme. We also still do simple Metropolis updates for $\theta$. Although this is not essential for ergodicity, it makes sampling a lot more efficient. 

\section{An empirical study}

The goal of our empirical study is to determine when using the novel updates improves sampling efficiency. We generate three simulated data sets that are representative of a range of possible scenarios, sample from the posterior distributions of $\theta$ for each of these data sets using various sampling schemes, and compute autocorrelation times to compare sampling efficiency.

\subsection{Simulated data from three scenarios}

The sort of data arising from the observation of a queueing system can be roughly classified into one of three scenarios. 

The first scenario is when the interarrival times are, on average, smaller than the service times, that is, arrivals are relatively frequent. In this case the queue is generally full, and empty only early on. Hence, most observed interdeparture times will tend to come from the service time distribution, and only a small number of interdeparture times will be relevant for inferring the arrival rate. Consequently, we expect there to be large uncertainty in $\theta_{3}$, but less for $\theta_{1}$ and $\theta_{2}$. 

The second scenario is when interarrival times are on average slightly larger than the service times. The queue is then sometimes empty, and sometimes contains a number of people. In this case, we expect the data to be informative about all parameters. This is also the case of most practical interest, as it corresponds to how we may expect a queueing system to behave in the real world.

The third scenario is when arrivals happen rarely, while the service time is relatively small. In this case, the queue will usually be empty. Interarrival times are then informative for inferring $\theta_{3}$, as most of them will come from the $\textnormal{Exp}(\theta_{3})$ distribution with a small amount of added $\textnormal{Uniform}(\theta_{1}, \theta_{2})$ noise. However, inference of the service time bounds $\theta_{1}$ and $\theta_{2}$ will now be based on how significantly the distribution of the interarrival times (for a given $\theta_{3}$) deviates from the Exponential distribution. This difference may be rather small, and so we expect that the posterior for the service time bounds will be rather diffuse.

The three data sets we generated each consisted of $n = 50$ interdeparture times, corresponding to each of the three scenarios above. For the first scenario we take $\theta = (8, 16, 0.15)$, for the second, $\theta = (4, 7, 0.15)$, and for the third, $\theta = (1, 2, 0.01)$.

\begin{figure}[t]
         \centering
         \begin{subfigure}[b]{0.3\textwidth}
                 \includegraphics[width=\textwidth]{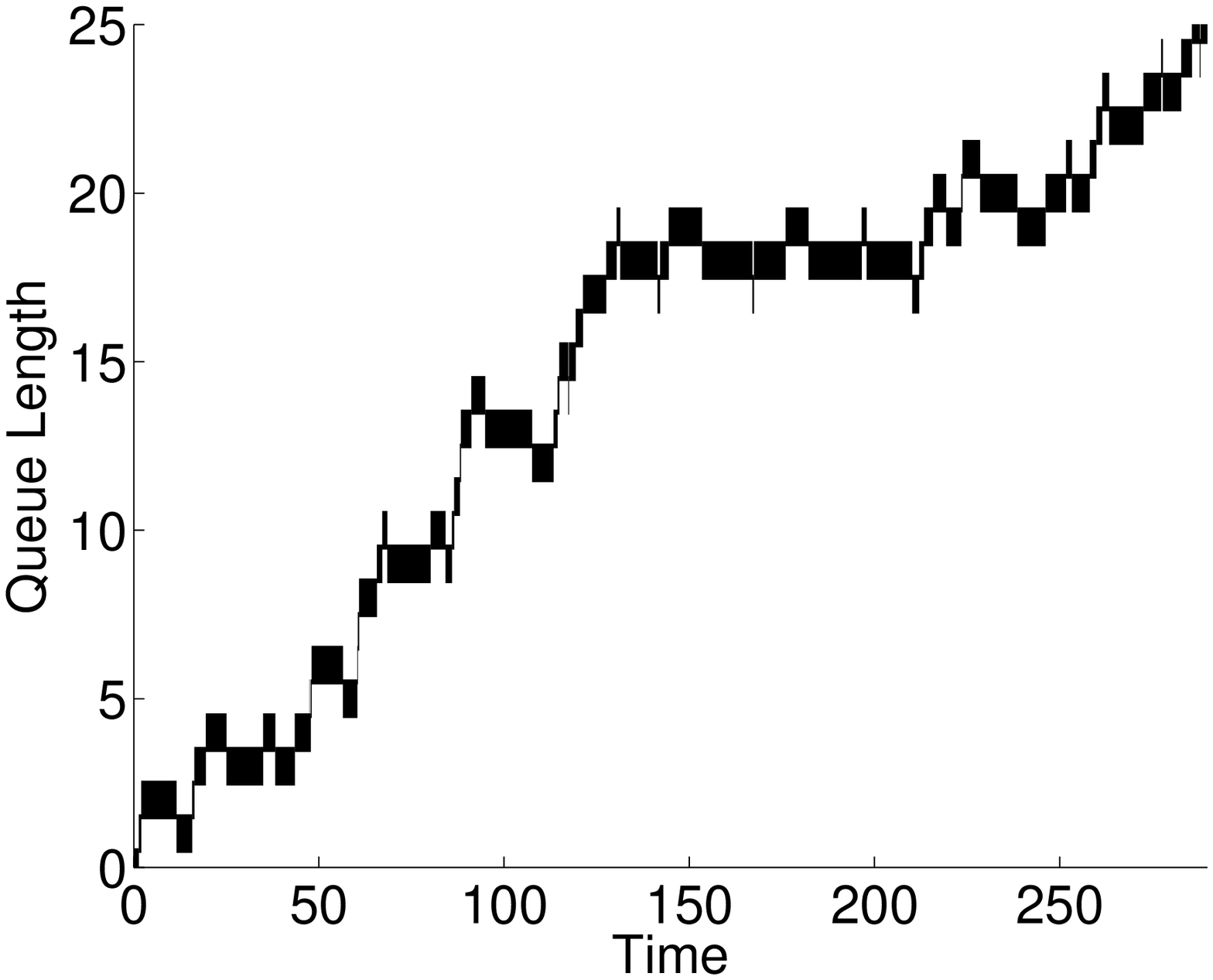}
                 \caption{$\theta = (8, 16, 0.15)$}
                 \label{fig:freq}
         \end{subfigure}
	 ~
         \begin{subfigure}[b]{0.3\textwidth}
                 \includegraphics[width=\textwidth]{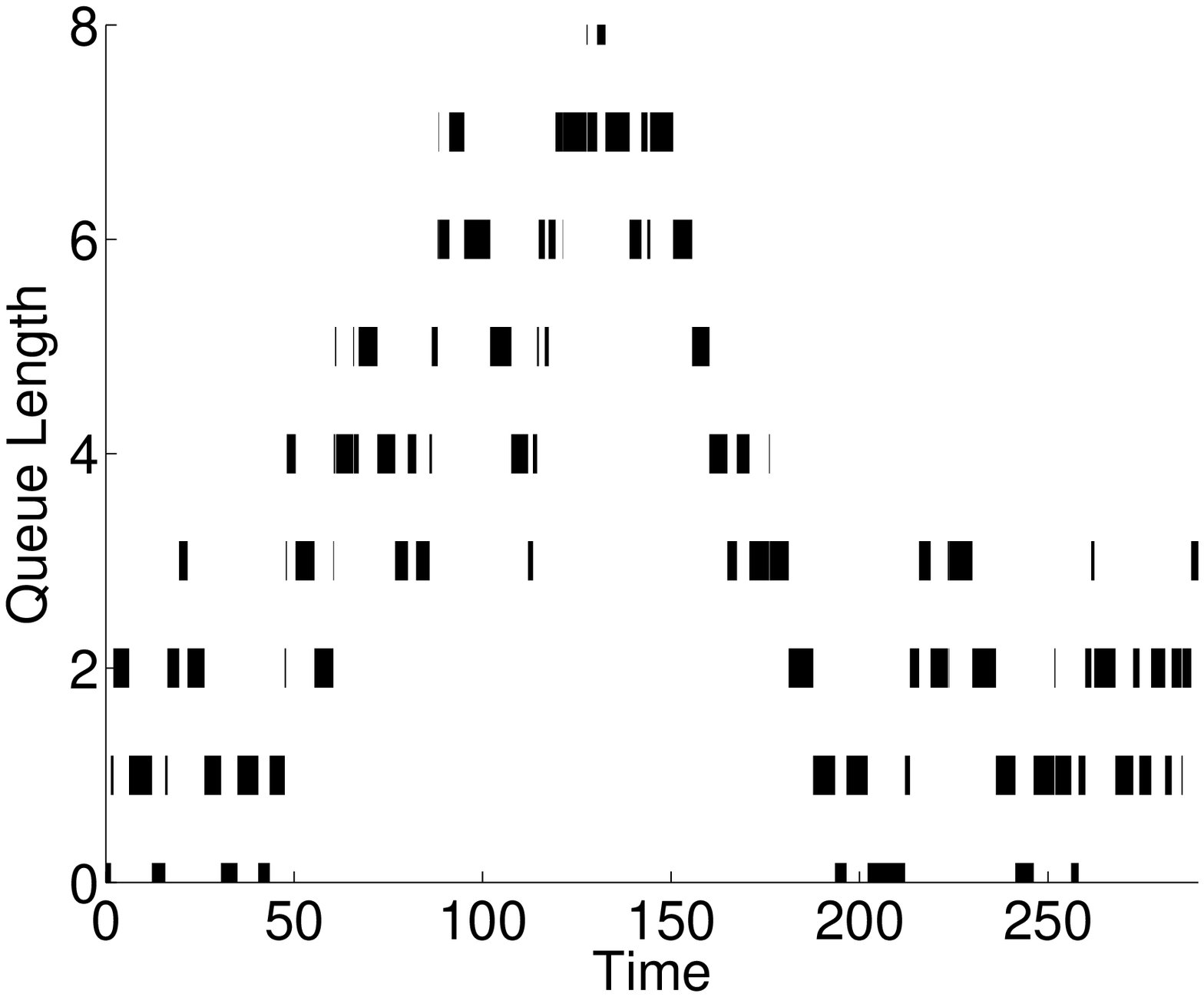}
                 \caption{$\theta = (4, 7, 0.15)$}
                 \label{fig:inter}
         \end{subfigure}
         ~
         \begin{subfigure}[b]{0.3\textwidth}
                 \includegraphics[width=\textwidth]{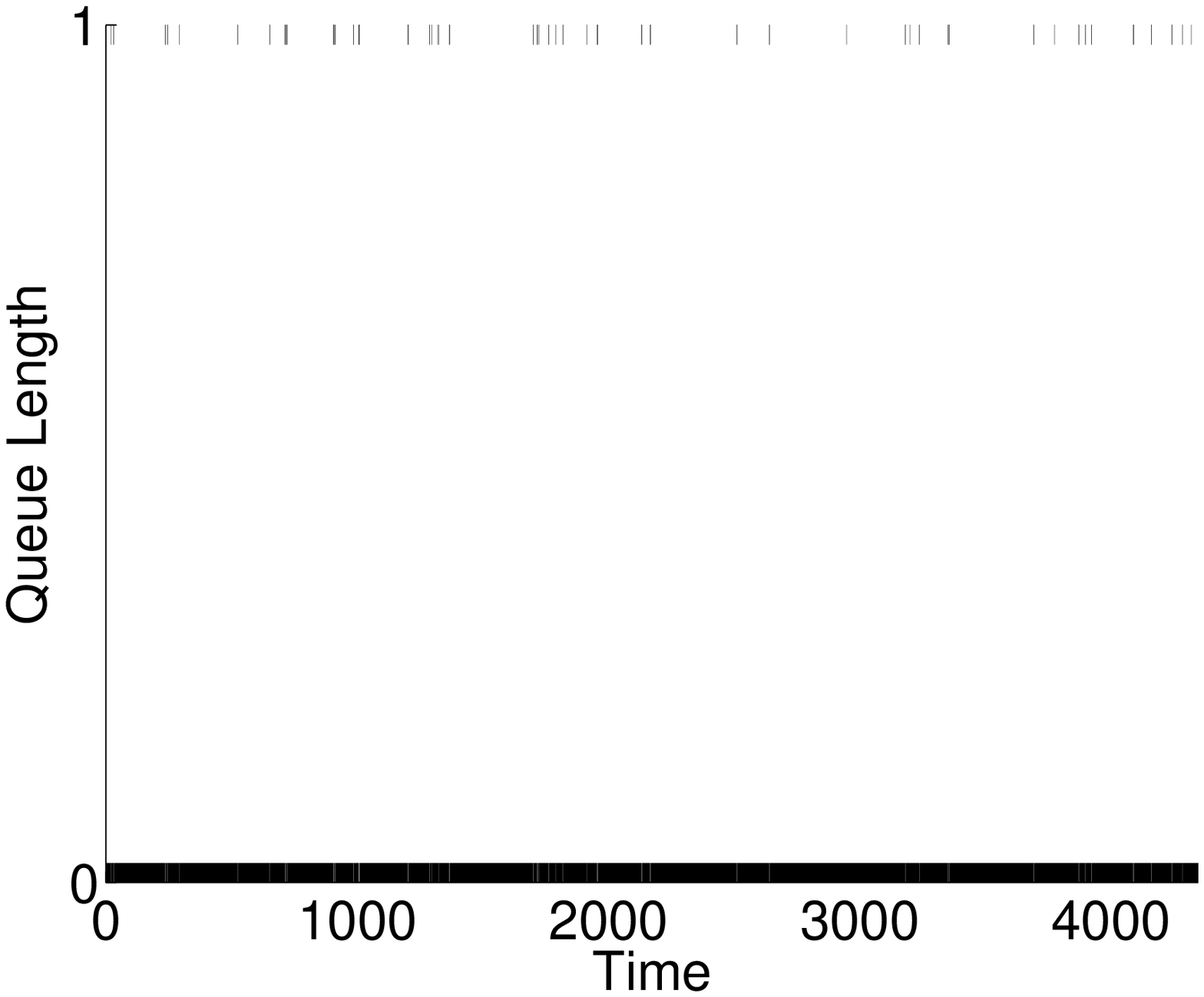}
                 \caption{$\theta = (1, 2, 0.01)$}
                 \label{fig:rare}
         \end{subfigure}
         \caption{The three data sets.}\label{fig:datasets}
\end{figure}

The plots in Figure \ref{fig:datasets} of the number of people in the queue up until the last arrival against time for each of the three data sets demonstrate that the data sets have the desired qualitative properties.

Numerical values of the simulated interdeparture times are presented in Table \ref{table:datasets}.

\begin{table}[p]
\tiny
\centering
\def\arraystretch{1.5}
\begin{tabular}{|r|r|r|}
\hline
Frequent & Intermediate & Rare  \\
\hline
11.57 &  6.19 & 21.77 \\ 
 13.44 &  6.04 & 10.30 \\ 
 13.24 &  9.52 & 206.34 \\ 
  9.30 &  4.49 &  8.57 \\ 
  8.95 &  4.36 & 45.79 \\ 
 11.99 &  9.86 & 233.13 \\ 
 15.68 &  9.91 & 128.30 \\ 
 10.72 &  5.02 & 59.73 \\ 
 12.68 &  5.76 &  4.59 \\ 
  9.79 &  4.67 &  3.21 \\ 
 14.01 &  6.25 & 185.29 \\ 
 10.04 &  4.77 &  2.49 \\ 
 12.05 &  5.52 &  4.63 \\ 
 13.59 &  6.10 & 72.48 \\ 
 15.13 &  6.67 & 22.47 \\ 
 15.67 &  6.88 & 195.34 \\ 
 12.38 &  5.64 & 85.92 \\ 
  9.11 &  4.42 &  8.39 \\ 
  9.19 &  4.45 & 23.30 \\ 
 10.06 &  4.77 &  4.24 \\ 
 14.73 &  6.52 & 42.78 \\ 
 10.03 &  4.76 & 332.64 \\ 
 14.51 &  6.44 & 16.91 \\ 
  9.95 &  4.73 &  6.26 \\ 
 15.43 &  6.79 & 39.44 \\ 
 10.80 &  5.05 & 27.16 \\ 
  9.57 &  4.59 & 29.53 \\ 
 10.01 &  4.75 & 93.65 \\ 
 12.93 &  5.85 & 42.60 \\ 
 11.79 &  5.42 & 176.36 \\ 
 10.81 &  5.05 & 34.69 \\ 
 14.65 &  6.49 & 345.20 \\ 
 12.68 &  5.76 & 128.16 \\ 
 12.40 &  8.67 & 307.50 \\ 
 15.34 & 16.65 & 233.54 \\ 
 10.29 &  4.86 & 18.79 \\ 
 14.06 &  6.27 & 36.88 \\ 
 14.03 &  6.26 & 114.85 \\ 
 11.04 &  5.14 &  4.73 \\ 
 12.54 & 10.60 & 337.02 \\ 
  8.61 &  4.23 & 81.89 \\ 
  8.43 &  6.15 & 96.33 \\ 
 12.25 &  5.59 & 27.20 \\ 
 14.23 &  6.34 & 23.16 \\ 
 15.47 &  6.80 & 167.89 \\ 
  9.04 &  4.39 & 70.58 \\ 
 12.55 &  5.71 & 81.28 \\ 
 11.76 &  5.41 & 43.55 \\ 
  8.10 &  4.04 & 33.88 \\ 
 10.70 &  5.01 & 28.47 \\ 
\hline
\end{tabular}
\caption{Simulated interdeparture times $y_{i}$}
\label{table:datasets}
\end{table}

\subsection{Experimental setup}

We now compare the efficiency of five different MCMC schemes for performing Bayesian inference for $\theta$ by drawing samples from the posterior. These are

1) The basic scheme: Gibbs sampling for $V_{i}$ with simple Metropolis updates for $\theta$.

2) Basic scheme plus shift updates. 

3) Basic scheme plus range scale updates.

4) Basic scheme plus rate scale updates. 

5) Basic scheme plus shift, rate scale, and range scale updates.

We put $\textnormal{Uniform}(0, 10)$ priors on $\theta_{1}$ and on $\theta_{2} - \theta_{1}$, and a $\textnormal{Uniform}(0, 1/3)$ prior on $\theta_{3}$. These are the priors that were used by Fearnhead and Prangle (2012). The prior $\pi(\theta)$ for $\theta$ is then
\begin{eqnarray}
\pi(\theta) &\propto& I(\theta_{1} \in [0, 10])I(\theta_{2} - \theta_{1} \in [0, 10])I(\theta_{3} \in [0, 1/3])
\end{eqnarray}
In the $\eta$ parametrization, the priors on $\eta_{1}$ and $\eta_{2}$ remain $\textnormal{Uniform}(0, 10)$, while the prior for $\eta_{3}$, due to the $\log$ transformation, is now proportional to $\exp(\eta_{3})$ on $(-\infty, \log(1/3))$ and $0$ otherwise. 

In order to compare MCMC methods fairly, we need to reasonably tune their parameters (such as the standard deviations of Metropolis proposals).  In practice, tuning is usually done by guesswork and trial and error.  But for these experiments, in order to avoid the influence of arbitrary choices, we use trial runs to ensure a reasonable choice of tuning parameters, even though in practice the time for such trial runs might exceed the gain compared to just making an educated guess.

We chose the standard deviations for the normal proposal in the simple Metropolis updates for $\eta$ by performing a number of pilot runs (using the basic scheme plus shift, rate scale, and range scale updates), finding estimates of the marginal posterior standard deviations of each component of $\eta$, and taking scalings of these estimated standard deviations as the corresponding proposal standard deviations.

The rationale for this choice of proposal standard deviations is as follows. One extreme case is when knowing the latent variables will give little additional information beyond the data for inferring the parameters, so the standard deviations of the marginal posterior will correspond closely to the standard deviations of the posterior given the latent variables and the data. The other extreme case is when the posterior given the latent variables and the data is a lot more concentrated than the marginal posterior, perhaps for a subset of the parameters. In most cases, however, the situation will be between these two extremes, so the marginal posterior standard deviations will provide a reasonable guide to setting proposal standard deviations. The case of rare arrivals is closer to the second extreme, when the knowing the latent variables will tend to strongly constrain $\eta_{1}$ and $\eta_{2}$, so we take a much smaller scaling for them than for $\eta_{3}$. 

For each simulated data set, we chose the number of simple Metropolis updates to perform in each iteration of a scheme by looking at the performance of the basic scheme with $1, 2, 4, 8, 16, 32$ Metropolis updates. 

For each shift update, we drew a shift $s$ from a $\textnormal{N}(0, \sigma_{\textnormal{shift}}^{2})$ distribution. For both scale updates we set fixed scales $c_{\textnormal{range}}$ and $c_{\textnormal{rate}}$. The chosen settings of tuning parameters for the different scenarios are presented in Table \ref{table:tuningset}.

We initialized $\eta_{1}$ to $\min(y_{i})$, $\eta_{2}$ and $\eta_{3}$ to their prior means, and the latent variables $V_{i}$ to $x_{i} - \min(y_{i})$, both for the pilot runs used to determine tuning settings and for the main runs used to compare the relative efficiency of different MCMC schemes.

\begin{table}[t]
\tiny
\centering
\def\arraystretch{1.5}
\begin{tabular}{|c|c|c|c|c|c|c|c|}
\hline
Scenario & Est. stdev. of $(\eta_{1}, \eta_{2}, \eta_{3})$ & Scaling  & Met. prop. stdev. & Met. updates & $\sigma_{\textnormal{shift}}^{2}$ & $c_{\textnormal{range}}$ & $c_{\textnormal{rate}}$ \\
\hline
Frequent & (0.1701, 0.2399, 0.3051) & 0.7 & (0.1191, 0.1679, 0.2136) & 1 & 0.3 & 1.008 & 1.7 \\
\hline
Intermediate & (0.0764, 0.1093, 0.1441) & 1 & (0.0764, 0.1093, 0.1441) & 16 & 0.2 & 1.03 & 1.004 \\
\hline
Rare & (0.6554, 2.0711, 0.1403) & (0.1, 0.1, 1)  & (0.0655,  0.2071,  0.1403) & 16 & 2 & 1.4 & 1.00005 \\
\hline
\end{tabular}
\caption{Tuning settings for the different samplers.}
\label{table:tuningset}
\end{table}

\subsection{Results}

We compared the peformance of our different MCMC schemes with the tuning settings in Table \ref{table:tuningset} on the three data scenarios. Run lengths were chosen to take about the same amount of time for each MCMC scheme, a total of about $43$ minutes per run using MATLAB on a Linux system with an Intel Xeon X5680 3.33 GHz CPU. Table \ref{table:runlengths} presents the run lengths. The acceptance rates of the various updates, across different data sets, were all in the range $17\%$ to $34\%$.

\begin{table}[t]
\tiny
\centering
\def\arraystretch{1.5}
\begin{tabular}{|c|c|c|c|c|c|}
\hline
Scenario & Basic & Basic + Shift & Basic + Range & Basic + Rate & Basic + All  \\
\hline
Frequent & 20 & 10.8 & 10.8 & 9.6 & 5.1 \\
\hline
Intermediate & 5.2 & 4.2 & 4.2 & 4 & 2.9 \\
\hline
Rare & 5.2 & 4.2 & 4.2 & 4 & 2.9 \\
\hline
\end{tabular}
\caption{Run lengths for different scenarios and samplers, in millions of iterations.}
\label{table:runlengths}
\end{table}

We first verified that the different sampling schemes give answers which agree by estimating the posterior means of the $\eta_{h}$, as well as the standard errors of the posterior mean estimates. For each combination of method and data set, we estimated the posterior means by taking a grand mean over the five MCMC runs, after discarding $10\%$ of each run as burn-in.  (In all cases, visual inspection of trace plots showed that the chain appears to have reached equilibrium after the first $10\%$ of iterations.) For each $\eta_{h}$, the standard errors of the posterior mean estimates were estimated by computing five posterior mean estimates using each of the five samples separately (discarding $10\%$ of each run as burn-in), computing the standard deviation of these posterior mean estimates, and dividing this standard deviation by $\sqrt{5}$. The approximate confidence intervals were obtained by taking the posterior mean estimate and adding or subtracting twice the estimated standard error of the posterior mean estimate. The results are shown in Table \ref{table:meanest}. There is no significant disagreement for posterior mean estimates across different methods.

\begin{table}[t]
	\tiny
	\begin{subtable}[b]{\textwidth}
		\centering
		\def\arraystretch{1.5}
		\begin{tabular}{|c|c|c|c|c|c|c|c|c|c|} 
		\hline 
		Parameter &\multicolumn{3}{c|}{$\eta_{1}$} & \multicolumn{3}{c|}{$\eta_{2}$} & \multicolumn{3}{c|}{$\eta_{3}$}\\ 
		\hline 
		Estimates & Mean & CI & std. err. & Mean & CI & std. err. & Mean & CI & std. err.\\ 
		\hline 
		Basic	 & 7.9292 & (7.9286, 7.9298) & 0.00031 & 7.9101 & (7.9092, 7.9110) & 0.00045 & -1.4817 & (-1.4853, -1.4781) & 0.00179 \\
		Basic + Shift	 & 7.9291 & (7.9287, 7.9296) & 0.00022 & 7.9102 & (7.9094, 7.9109) & 0.00038 & -1.4774 & (-1.4816, -1.4733) & 0.00207 \\
		Basic + Range	 & 7.9292 & (7.9288, 7.9296) & 0.00019 & 7.9102 & (7.9098, 7.9107) & 0.00024 & -1.4778 & (-1.4796, -1.4760) & 0.00090 \\
		Basic + Rate	 & 7.9292 & (7.9287, 7.9296) & 0.00024 & 7.9102 & (7.9094, 7.9110) & 0.00041 & -1.4835 & (-1.4837, -1.4833) & 0.00012 \\
		Basic + All	 & 7.9293 & (7.9286, 7.9301) & 0.00037 & 7.9100 & (7.9087, 7.9112) & 0.00063 & -1.4834 & (-1.4836, -1.4832) & 0.00011 \\
		\hline 
		\end{tabular}
		\caption{Frequent arrivals.}
	        \vspace{16pt}
	\end{subtable}
	~
	\begin{subtable}[b]{\textwidth}
		\centering
		\def\arraystretch{1.5}
		\begin{tabular}{|c|c|c|c|c|c|c|c|c|c|} 
		\hline 
		Parameter &\multicolumn{3}{c|}{$\eta_{1}$} & \multicolumn{3}{c|}{$\eta_{2}$} & \multicolumn{3}{c|}{$\eta_{3}$}\\ 
		\hline 
		Estimates & Mean & CI & std. err. & Mean & CI & std. err. & Mean & CI & std. err.\\ 
		\hline 
		Basic	 & 3.9612 & (3.9611, 3.9613) & 0.00004 & 2.9866 & (2.9865, 2.9867) & 0.00006 & -1.7317 & (-1.7318, -1.7317) & 0.00002 \\
		Basic + Shift	 & 3.9611 & (3.9610, 3.9612) & 0.00004 & 2.9866 & (2.9865, 2.9868) & 0.00007 & -1.7317 & (-1.7318, -1.7316) & 0.00006 \\
		Basic + Range	 & 3.9612 & (3.9611, 3.9613) & 0.00004 & 2.9865 & (2.9864, 2.9866) & 0.00005 & -1.7318 & (-1.7318, -1.7317) & 0.00004 \\
		Basic + Rate	 & 3.9611 & (3.9610, 3.9613) & 0.00007 & 2.9866 & (2.9864, 2.9868) & 0.00010 & -1.7317 & (-1.7318, -1.7316) & 0.00004 \\
		Basic + All	 & 3.9612 & (3.9611, 3.9612) & 0.00003 & 2.9865 & (2.9864, 2.9866) & 0.00006 & -1.7316 & (-1.7317, -1.7316) & 0.00003 \\
		\hline 
		\end{tabular}
		\caption{Intermediate case.}
	        \vspace{16pt}
	\end{subtable}

	\begin{subtable}[b]{\textwidth}
		\centering
		\def\arraystretch{1.5}
		\begin{tabular}{|c|c|c|c|c|c|c|c|c|c|} 
		\hline 
		Parameter &\multicolumn{3}{c|}{$\eta_{1}$} & \multicolumn{3}{c|}{$\eta_{2}$} & \multicolumn{3}{c|}{$\eta_{3}$}\\ 
		\hline 
		Estimates & Mean & CI & std. err. & Mean & CI & std. err. & Mean & CI & std. err.\\ 
		\hline 
		Basic	 & 1.6986 & (1.6878, 1.7094) & 0.00538 & 4.2875 & (4.2330, 4.3420) & 0.02725 & -4.4549 & (-4.4551, -4.4546) & 0.00013 \\
		Basic + Shift	 & 1.7012 & (1.6984, 1.7039) & 0.00138 & 4.3098 & (4.2667, 4.3529) & 0.02154 & -4.4549 & (-4.4550, -4.4548) & 0.00005 \\
		Basic + Range	 & 1.7038 & (1.7002, 1.7074) & 0.00181 & 4.2737 & (4.2632, 4.2841) & 0.00520 & -4.4549 & (-4.4551, -4.4547) & 0.00010 \\
		Basic + Rate	 & 1.7018 & (1.6953, 1.7083) & 0.00325 & 4.3077 & (4.2631, 4.3523) & 0.02230 & -4.4549 & (-4.4550, -4.4548) & 0.00006 \\
		Basic + All	 & 1.7003 & (1.6996, 1.7010) & 0.00036 & 4.2846 & (4.2751, 4.2942) & 0.00477 & -4.4549 & (-4.4552, -4.4546) & 0.00013 \\
		\hline 
		\end{tabular}
		\caption{Rare arrivals.}
	\end{subtable}

\caption{Posterior mean estimates with approximate CI's and standard errors}
\label{table:meanest}
\end{table}

Having established that the different methods we want to compare give answers which agree, we next compare their efficiency by looking at the autocorrelation times, $\tau_{h}$, of the Markov chains used to sample $\eta_{h}$. The autocorrelation time is a common MCMC performance metric that can be roughly interpreted as the number of draws one needs to make with the MCMC sampler to get the equivalent of one independent point (Neal (1993)) and is defined as
\begin{eqnarray}
\tau_{h} &=& 1 + 2 \sum_{i=1}^{\infty} \rho_{h, k}
\end{eqnarray}
where $\rho_{h, k}$ is the autocorrelation at lag $k$ of the chain used to sample $\eta_{h}$. 

For each combination of method and data set, for each $\eta_{h}$, we estimate $\tau_{h}$ by
\begin{eqnarray}
\hat{\tau}_{h} &=& 1 + 2 \sum_{i=1}^{K_{h}} \hat{\rho}_{h, k}
\end{eqnarray}
where the truncation point $K_{h}$ is such that for $k > K_{h}$, the estimate $\hat{\rho}_{h, k}$ is not appreciably different from $0$.

To estimate $\rho_{h, k}$, we use estimates of the lag $k$ autocovariances $\gamma_{h, k}$ for $k = 0, \ldots, K_{h}$. These are obtained as follows. For each of the five samples, indexed by $s = 1, \ldots 5$ and drawn using some method, we first compute an autocovariance estimate
\begin{eqnarray}
\hat{\gamma}_{h, k}^{s} &=& \frac{1}{M}\sum_{m = 1}^{M - k}(\eta_{h}^{[m, s]} - \overline{\eta_{h}})(\eta_{h}^{[m+k, s]} - \overline{\eta_{h}})
\end{eqnarray}
Here $M$ the length of the sample (after discarding $10\%$ of the run as burn-in) and $\eta_{h}^{[m, s]}$ the $m$-th value of $\eta_{h}$ in the $s$-th sample. We take $\overline{\eta_{h}}$ to be the grand mean of $\eta_{h}$ over all five samples (each of these samples has length $M$). We do this because it allows us to detect if the Markov chain explores different regions of the parameter and latent variable space for different random number generator seeds. (When the mean from one or more of the five samples differs substantially from the grand mean, autocovariance estimates will be much higher.)

We then estimate $\gamma_{h, k}$ by averaging autocovariance estimates from the five samples
\begin{eqnarray}
\hat{\gamma}_{h, k} &=& \frac{1}{5} \sum_{s = 1}^{5} \hat{\gamma}_{h, k}^{s}
\end{eqnarray}
and we estimate $\rho_{h, k}$ for $k = 1, \ldots, K_{h}$ with $\hat{\gamma}_{h, k} / \hat{\gamma}_{h, 0}$. (In practice, for long runs, it is much more efficient to use the fast Fourier transform (FFT) to compute the autocovariance estimates.) Table \ref{table:tauest} shows the estimated autocorrelation times for different sampling schemes. To compare the methods fairly, we multiply each autocorrelation time by the average time per iteration.

\begin{table}[t]
\tiny
\centering
\def\arraystretch{1.5}
\setlength{\tabcolsep}{0.15cm}
\begin{tabular}{|c|c|c|c|c|c|c|c|c|c|c|c|c|c|c|c|c|c|c|c|} 
\hline 
Scenario & \multicolumn{3}{c|}{Frequent} & \multicolumn{3}{c|}{Intermediate} & \multicolumn{3}{c|}{Rare} & Time (ms) & \multicolumn{3}{c|}{Frequent} & \multicolumn{3}{c|}{Intermediate} & \multicolumn{3}{c|}{Rare} \\ 
\hline 
Parameter &$\eta_{1}$ & $\eta_{2}$ & $\eta_{3}$ &$\eta_{1}$ & $\eta_{2}$ & $\eta_{3}$ &$\eta_{1}$ & $\eta_{2}$ & $\eta_{3}$ &Freq./Inter./Rare &$\eta_{1}$ & $\eta_{2}$ & $\eta_{3}$ &$\eta_{1}$ & $\eta_{2}$ & $\eta_{3}$ &$\eta_{1}$ & $\eta_{2}$ & $\eta_{3}$\\ 
\hline 
Basic	 &       99 &       98 &     7800 &      5.4 &      6.1 &      3.2 &      1400 &     4400 &      5.6 &      0.13/0.50/0.50 &       13 &       13 &     1000 &      2.7 &        3 &      1.6 &      700 &     2200 &        2.8 \\
Basic + Shift	 &       46 &       75 &     7400 &      4.4 &      5.8 &      3.2 &      130 &     2100 &      4.1 &     0.24/0.61/0.61 &       11 &       18 &     1800 &      2.7 &      3.5 &        2.0 &       79 &     1300 &      2.5 \\
Basic + Range	 &       95 &       86 &     5200 &      5.2 &      5.3 &      3.2 &      380 &       73 &      4.6 &     0.24/0.62/0.62 &       23 &       21 &     1200 &      3.2 &      3.3 &        2.0 &      240 &       45 &      2.9 \\
Basic + Rate	 &       95 &       94 &       13 &      5.4 &        6.0 &      3.2 &     1100 &     2800 &      4.5 &     0.27/0.65/0.65 &       26 &       25 &      3.5 &      3.5 &      3.9 &      2.1 &     720 &     1800 &      2.9 \\
Basic + All	 &       36 &       55 &       11 &      4.2 &        5.0 &      3.2 &       13 &       40 &      4.2 &     0.51/0.89/0.89 &       18 &       28 &      5.6 &      3.7 &      4.5 &      2.8 &       12 &       36 &      3.7 \\
\hline 
\end{tabular}
\caption{Estimates of autocorrelation times for different methods. Unadjusted autocorrelation times are on the left, autocorrelation times multiplied by the average time per iteration are on the right.}
\label{table:tauest}
\end{table}

From Table \ref{table:tauest}, we see that using just the shift, or just a scale update (either a range or a rate scale update) improves performance for sampling parameters that are changed by one of these updates. For a scheme which uses all updates, the greatest improvement in performance for sampling $\eta_{3}$ is in the case of frequent arrivals (efficiency gain of 179 times), while perfomance improvement when sampling $\eta_{1}$ and $\eta_{2}$ is greatest in the case of rare arrivals (efficiency gains of 58 and 61 times). In other cases, performance neither increases nor decreases significantly. These results are in approximate agreement with the estimates of the standard errors of the posterior mean estimates in Table \ref{table:meanest}.

In an additional run (not used to compute the autocorrelation times shown here) of the basic plus rate scheme, for the rare arrivals scenario, we found that the sampler got stuck for a while in a region of the parameter space. The basic and basic + shift methods (when initialized to a state from this ``stuck'' region) also stayed in this ``stuck'' region for a while before visting other regions. The basic plus range and basic plus all methods, when initialized with the stuck state, quickly returned to sampling other regions. So, autocorrelation time estimates for the rare arrivals scenario, for methods other than basic plus all and basic plus rate, probably underestimate actual autocorrelation times.

\begin{figure}[p]
         \centering
         \begin{subfigure}[b]{0.45\textwidth}
                 \includegraphics[width=\textwidth]{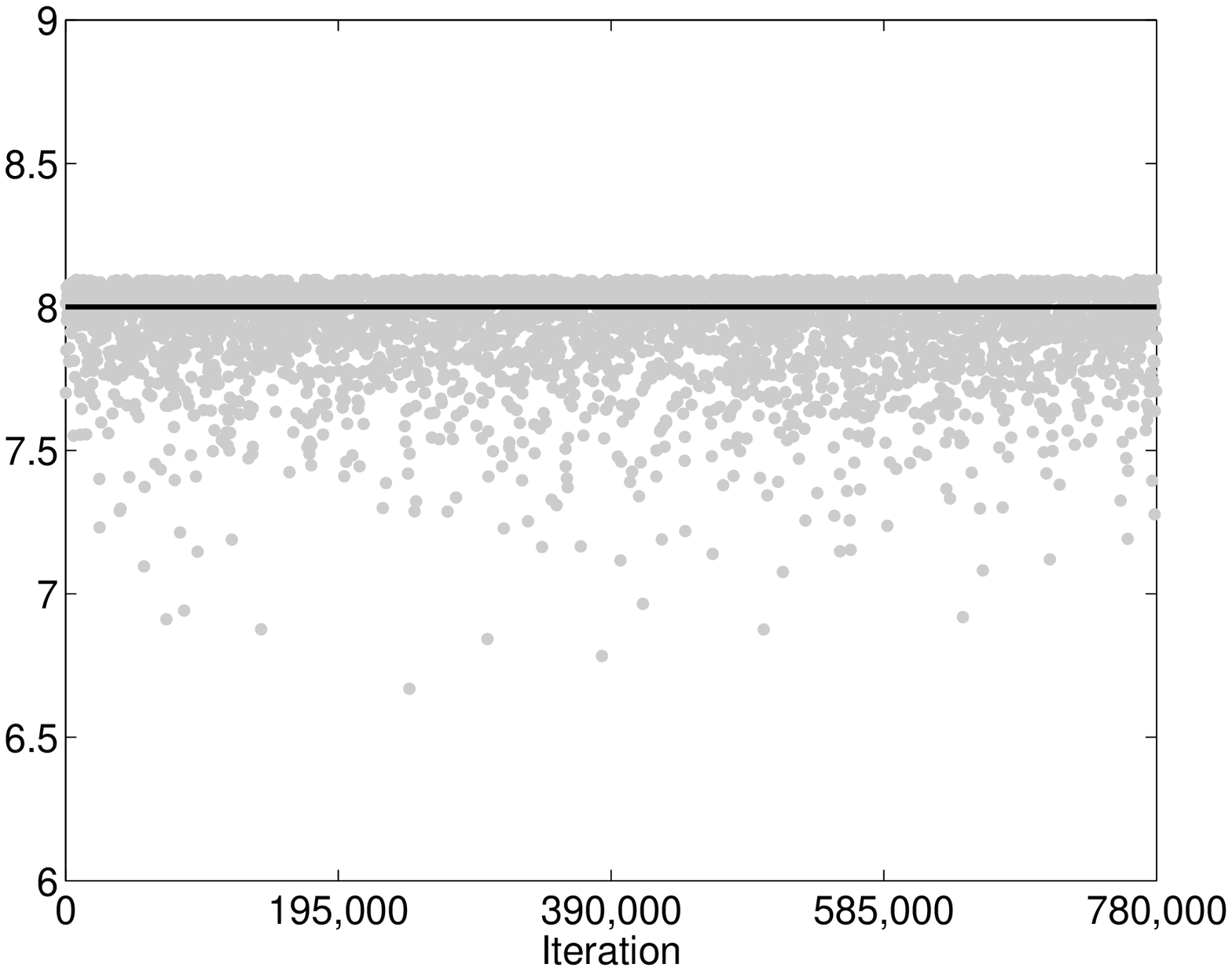}
                 \caption{$\eta_{1}$, Basic scheme.}
		 \vspace{12pt}
         \end{subfigure}
	~
         \begin{subfigure}[b]{0.45\textwidth}
                 \includegraphics[width=\textwidth]{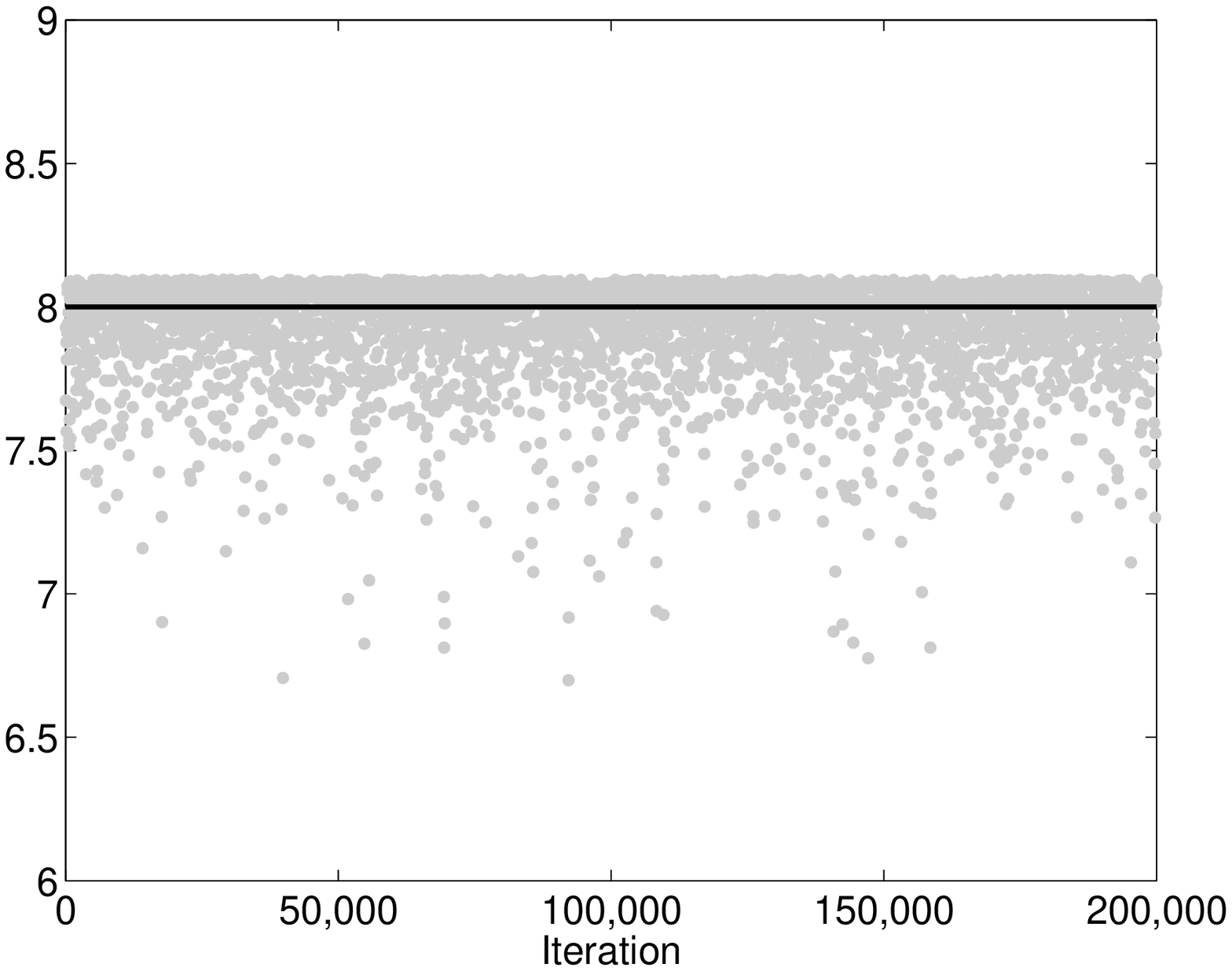}
                 \caption{$\eta_{1}$, Basic + all.}
		 \vspace{12pt}
         \end{subfigure}

         \centering
         \begin{subfigure}[b]{0.45\textwidth}
                 \includegraphics[width=\textwidth]{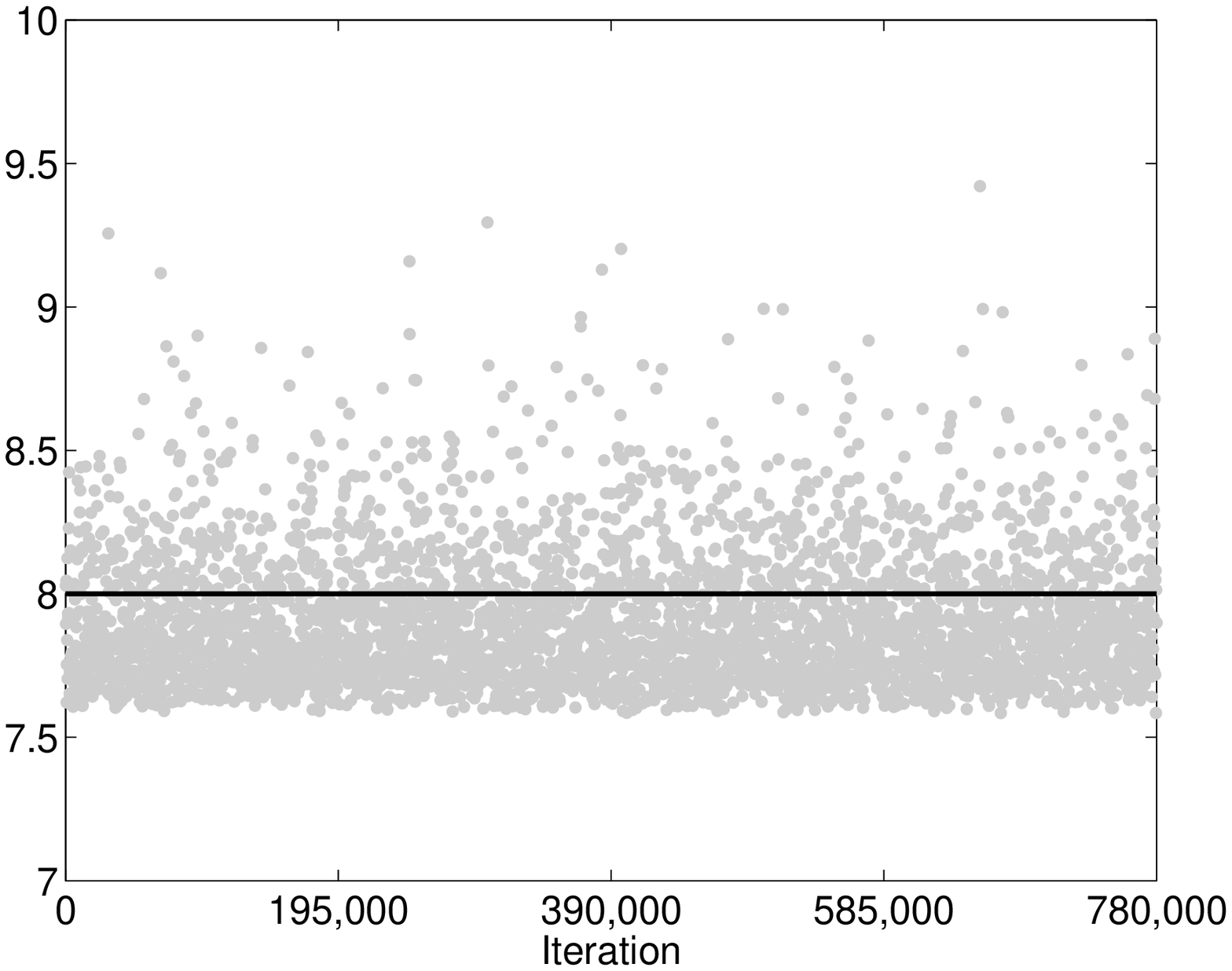}
                 \caption{$\eta_{2}$, Basic scheme.}
		 \vspace{12pt}
         \end{subfigure}
	~
         \begin{subfigure}[b]{0.45\textwidth}
                 \includegraphics[width=\textwidth]{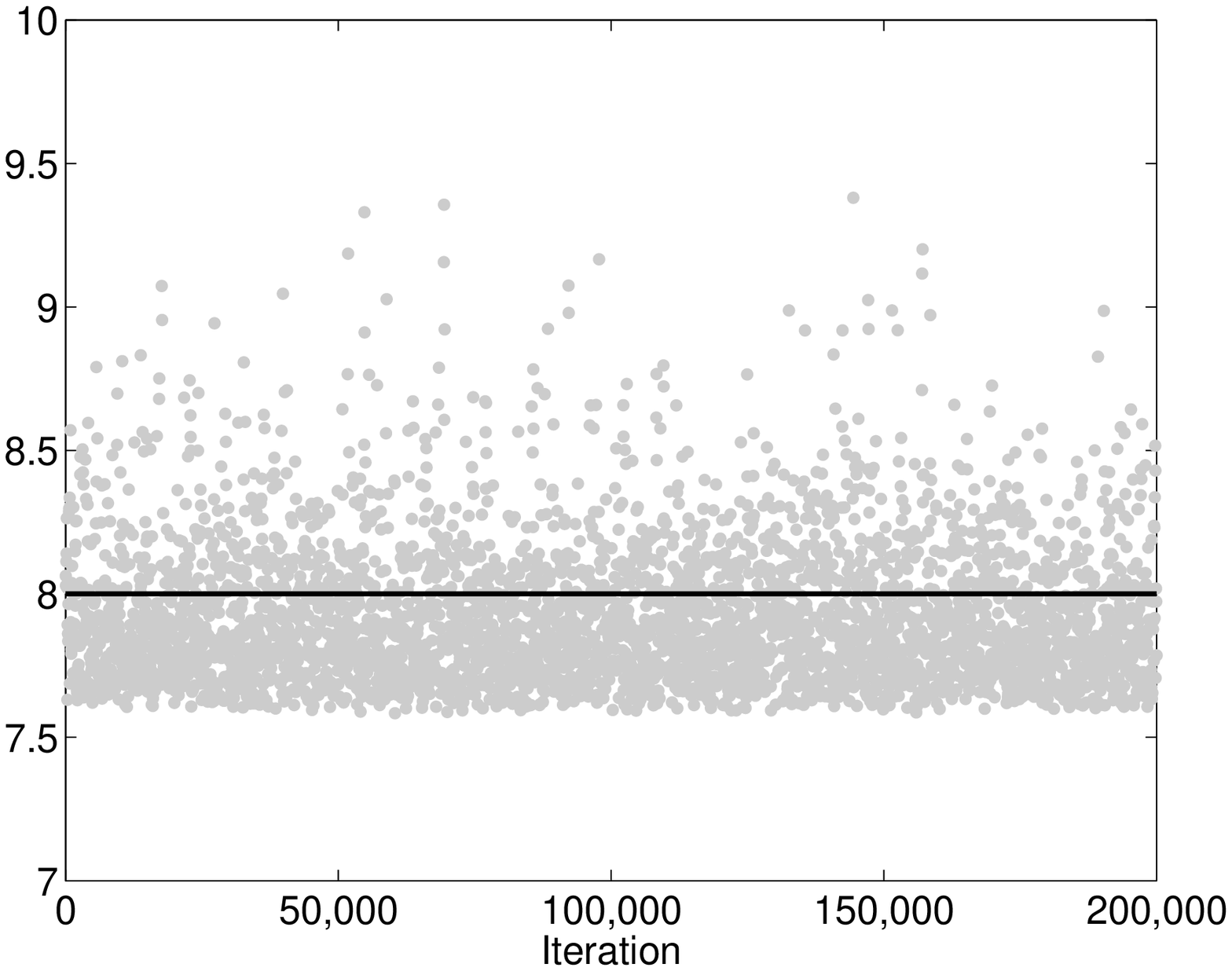}
                 \caption{$\eta_{2}$, Basic + all.}
		 \vspace{12pt}
         \end{subfigure}

         \centering
         \begin{subfigure}[b]{0.45\textwidth}
                 \includegraphics[width=\textwidth]{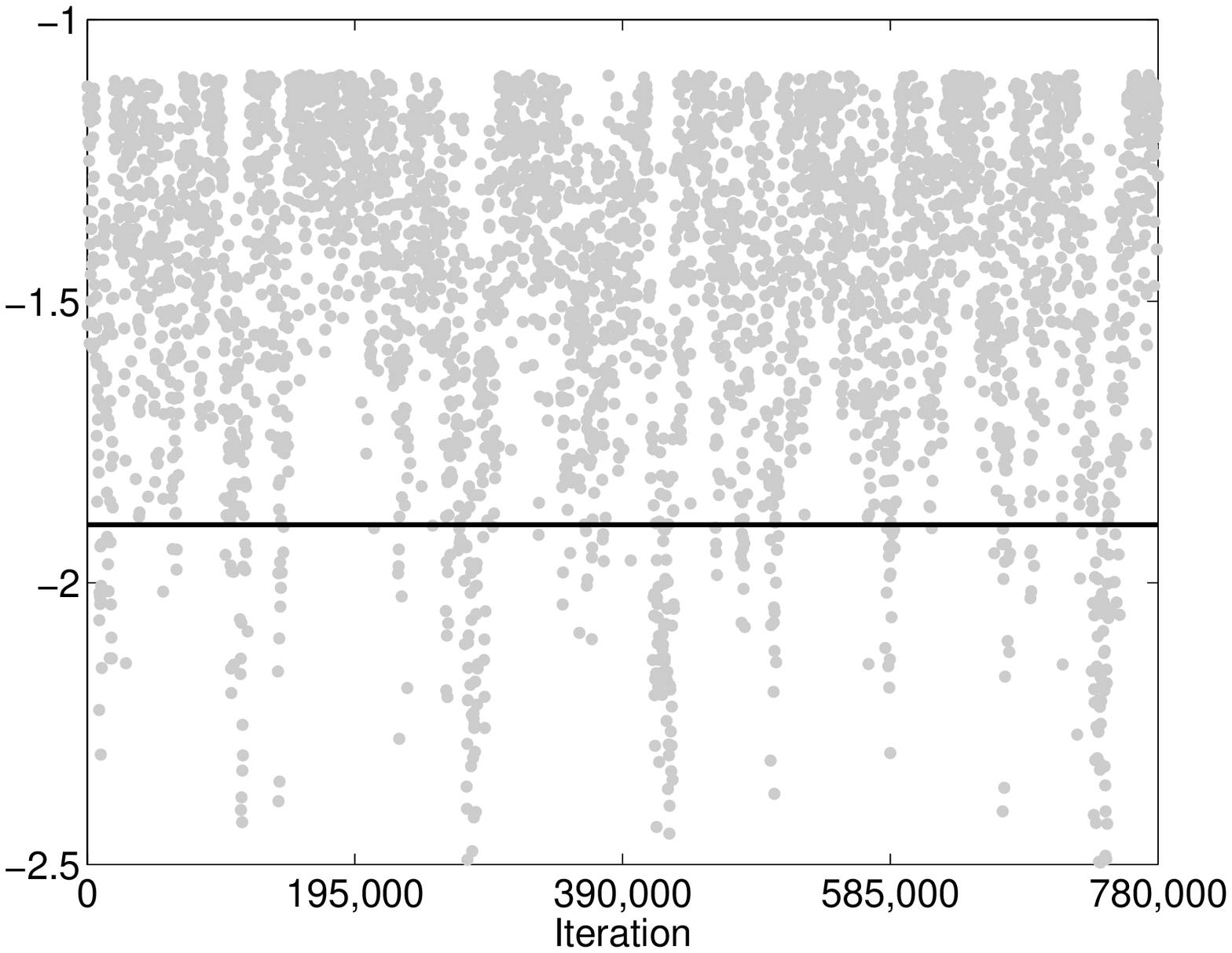}
                 \caption{$\eta_{3}$, Basic scheme.}
		 \vspace{12pt}
         \end{subfigure}
	~
         \begin{subfigure}[b]{0.45\textwidth}
                 \includegraphics[width=\textwidth]{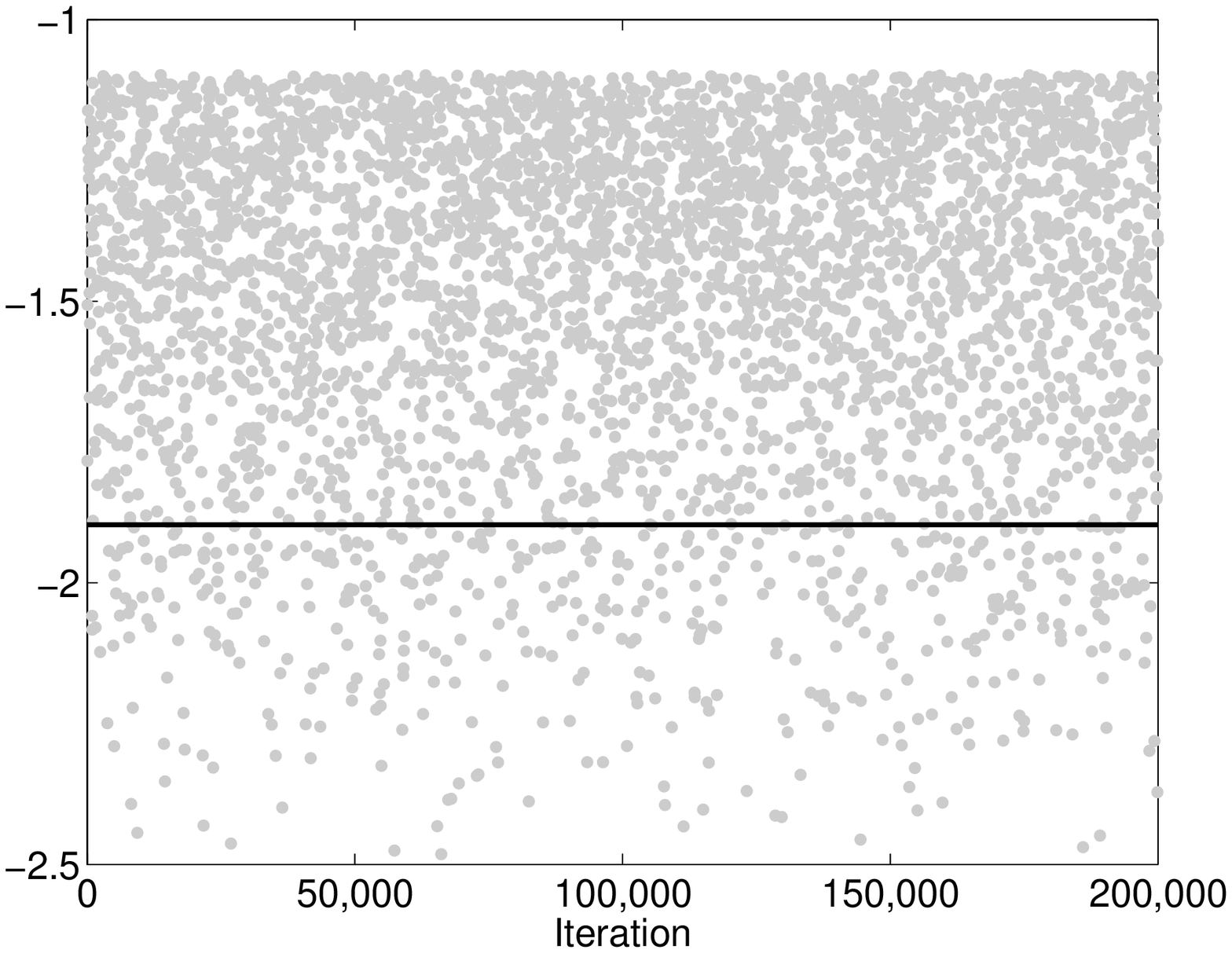}
                 \caption{$\eta_{3}$, Basic + all.}
		 \vspace{12pt}
         \end{subfigure}

         \caption{Comparison of performance for frequent arrivals.}\label{fig:tracefreq}

\end{figure}

\begin{figure}[p]
         \centering
         \begin{subfigure}[b]{0.45\textwidth}
                 \includegraphics[width=\textwidth]{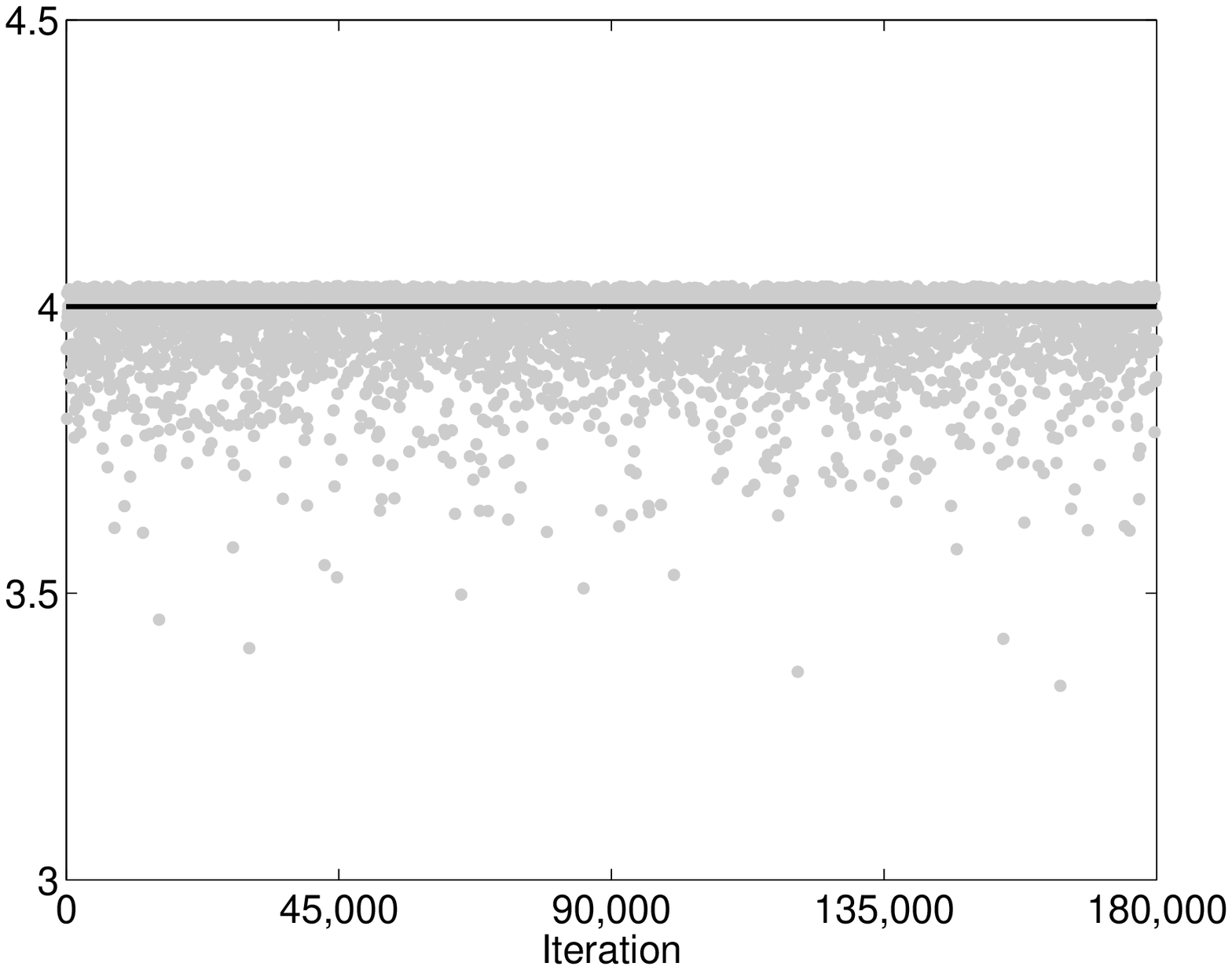}
                 \caption{$\eta_{1}$, Basic scheme.}
		 \vspace{12pt}
         \end{subfigure}
	~
         \begin{subfigure}[b]{0.45\textwidth}
                 \includegraphics[width=\textwidth]{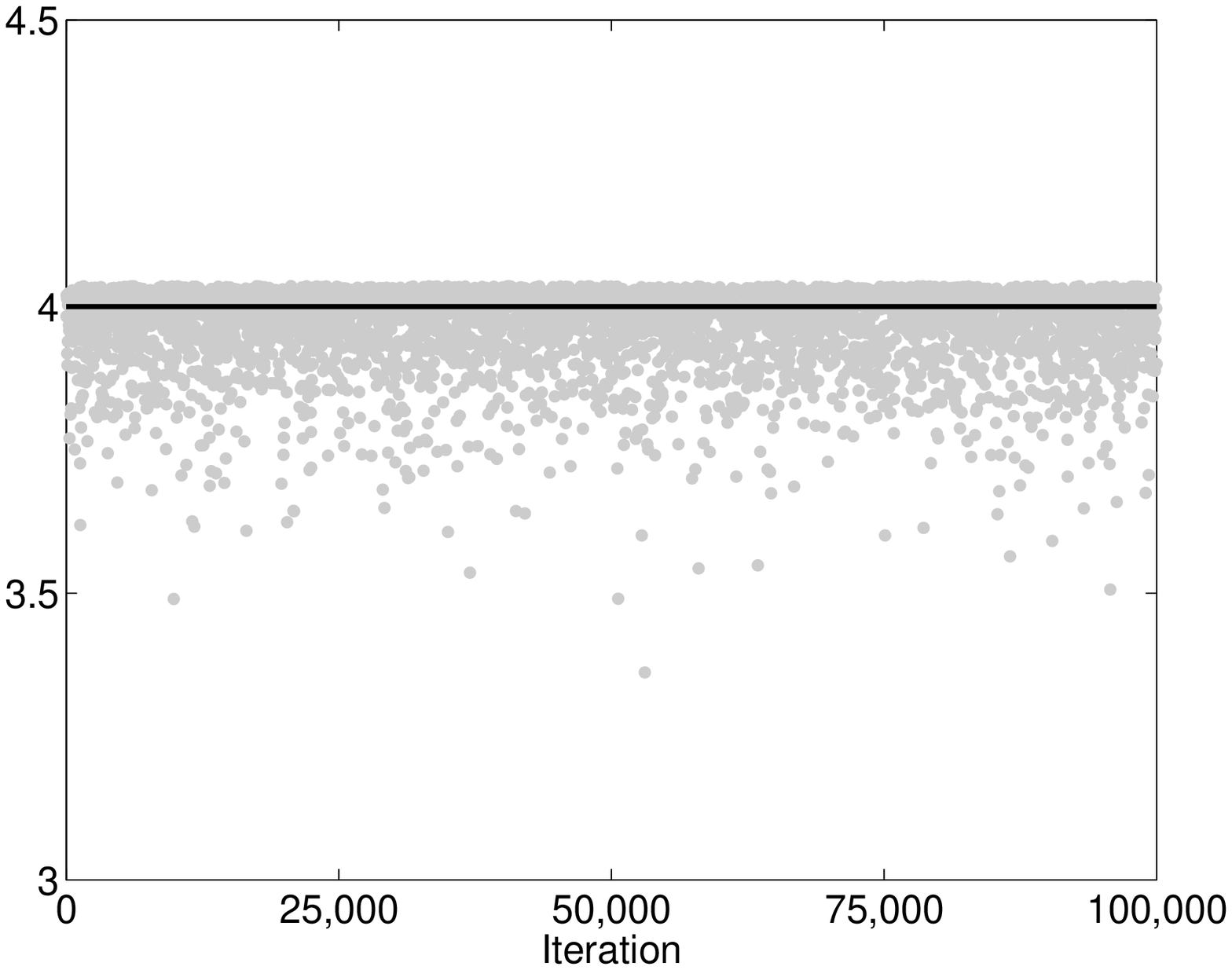}
                 \caption{$\eta_{1}$, Basic + all.}
		 \vspace{12pt}
         \end{subfigure}

         \centering
         \begin{subfigure}[b]{0.45\textwidth}
                 \includegraphics[width=\textwidth]{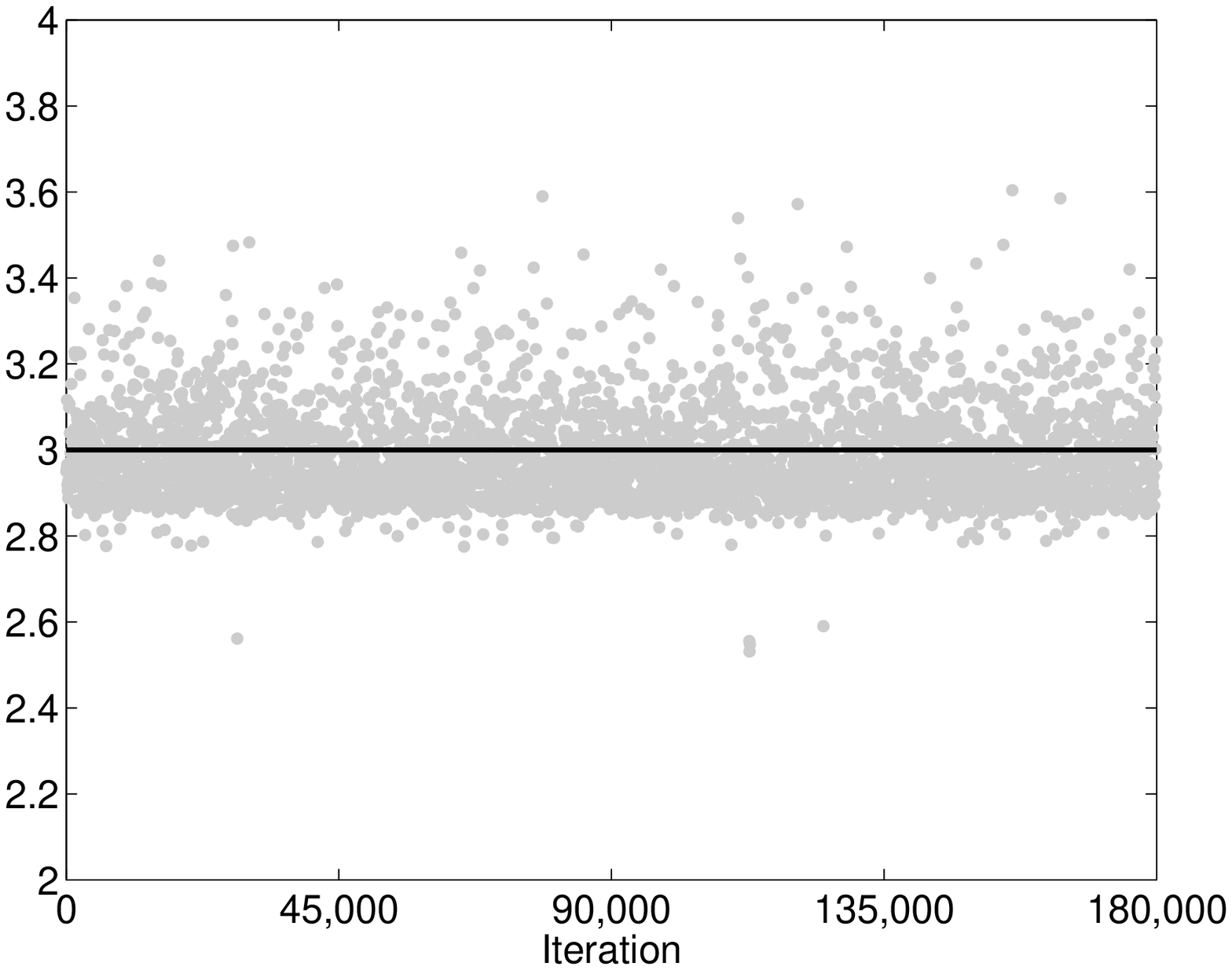}
                 \caption{$\eta_{2}$, Basic scheme.}
		 \vspace{12pt}
         \end{subfigure}
	~
         \begin{subfigure}[b]{0.45\textwidth}
                 \includegraphics[width=\textwidth]{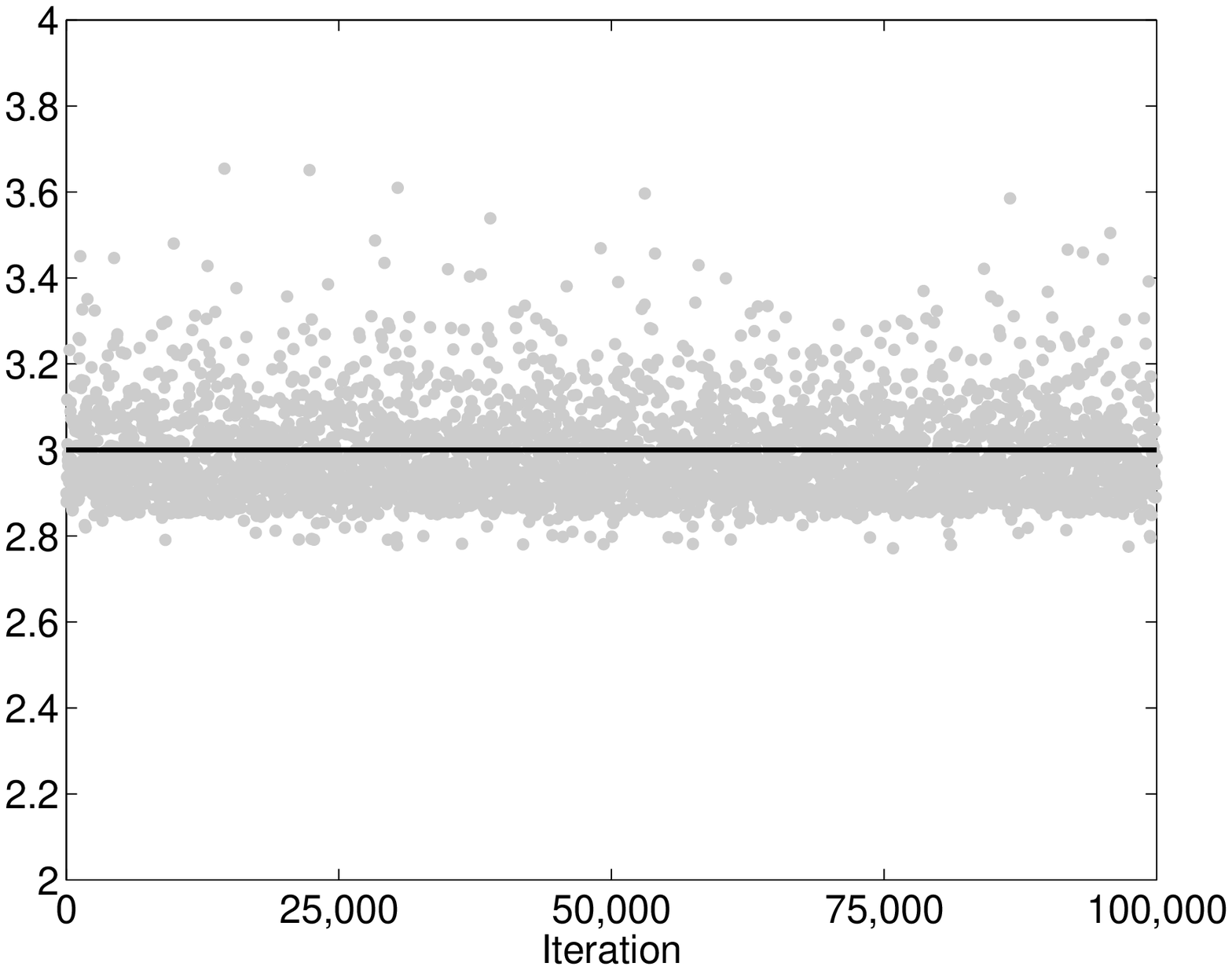}
                 \caption{$\eta_{2}$, Basic + all.}
		 \vspace{12pt}
         \end{subfigure}

         \centering
         \begin{subfigure}[b]{0.45\textwidth}
                 \includegraphics[width=\textwidth]{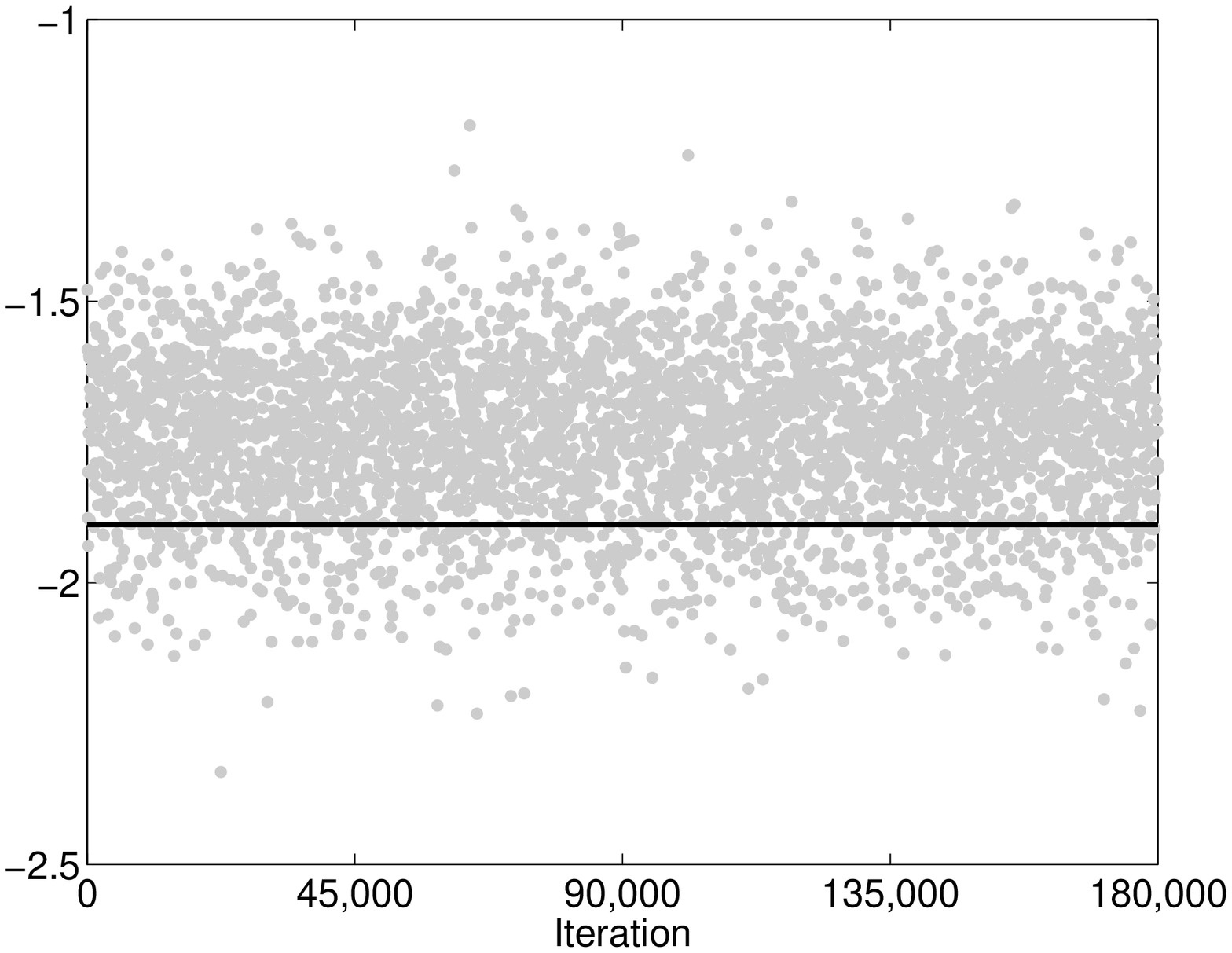}
                 \caption{$\eta_{3}$, Basic scheme.}
		 \vspace{12pt}
         \end{subfigure}
	~
         \begin{subfigure}[b]{0.45\textwidth}
                 \includegraphics[width=\textwidth]{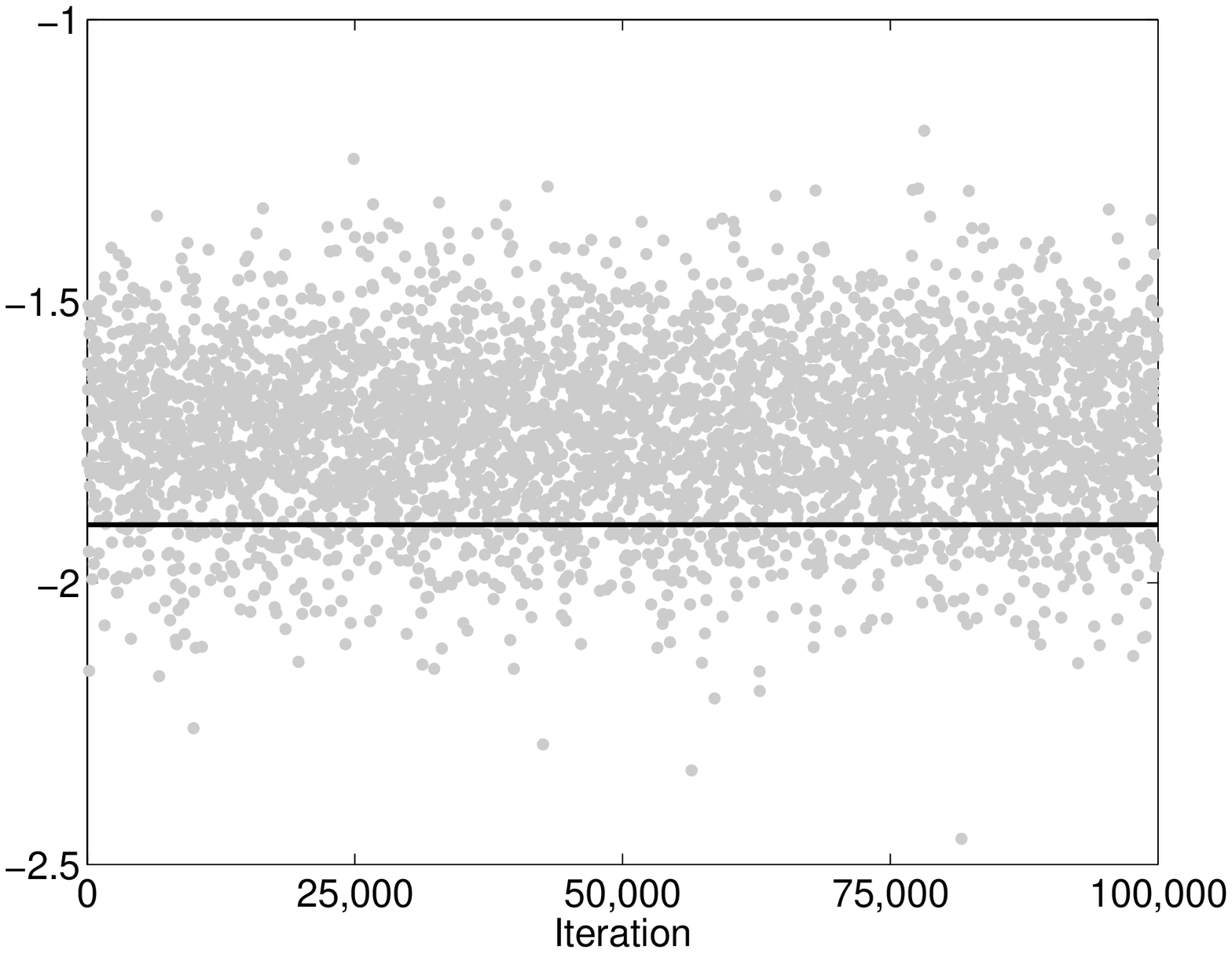}
                 \caption{$\eta_{3}$, Basic + all.}
		 \vspace{12pt}
         \end{subfigure}
         \caption{Comparison of performance for intermediate case.}\label{fig:traceinter}

\end{figure}

\begin{figure}[p]
         \centering
         \begin{subfigure}[b]{0.45\textwidth}
                 \includegraphics[width=\textwidth]{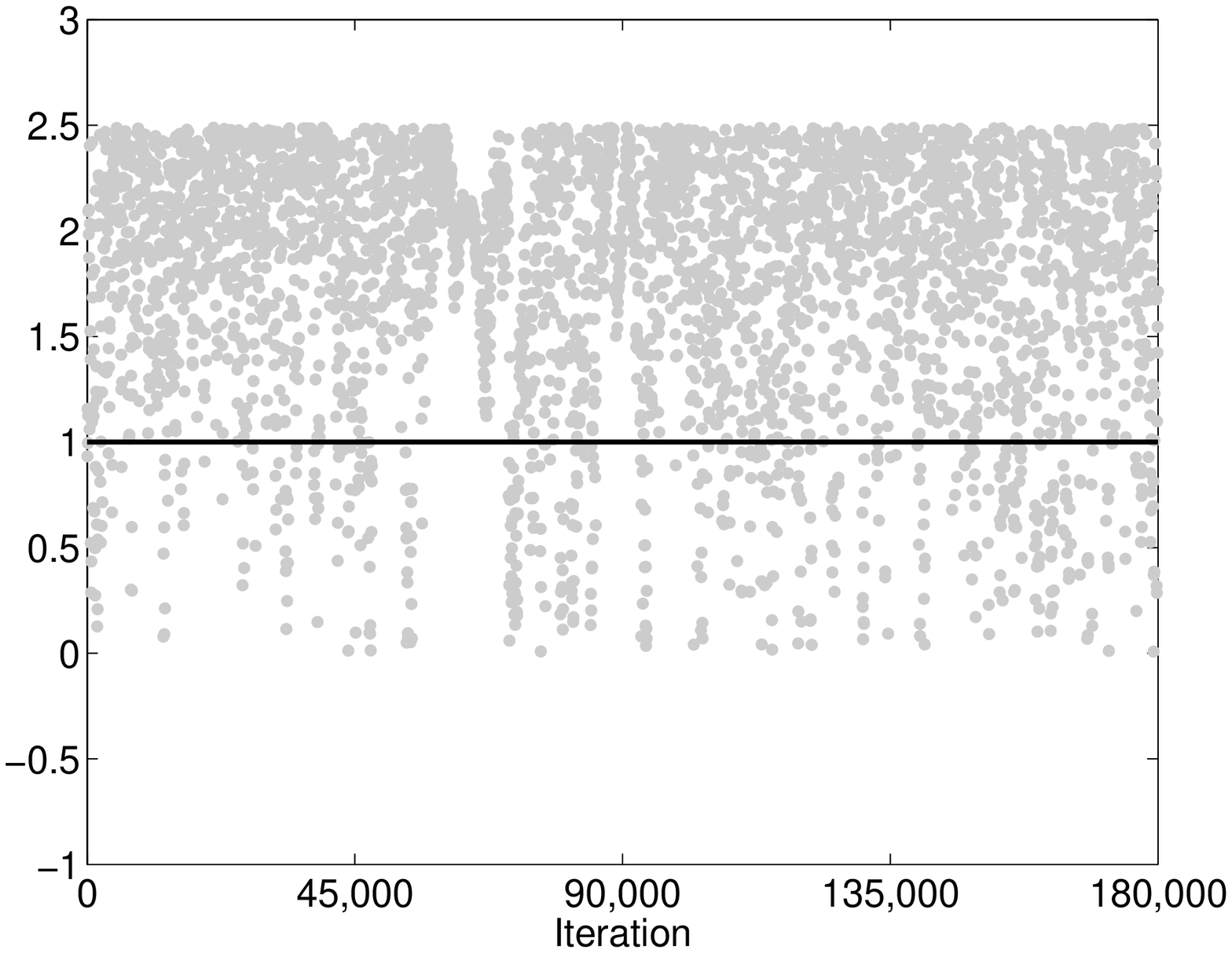}
                 \caption{$\eta_{1}$, Basic scheme.}
		 \vspace{12pt}
         \end{subfigure}
	~
         \begin{subfigure}[b]{0.45\textwidth}
                 \includegraphics[width=\textwidth]{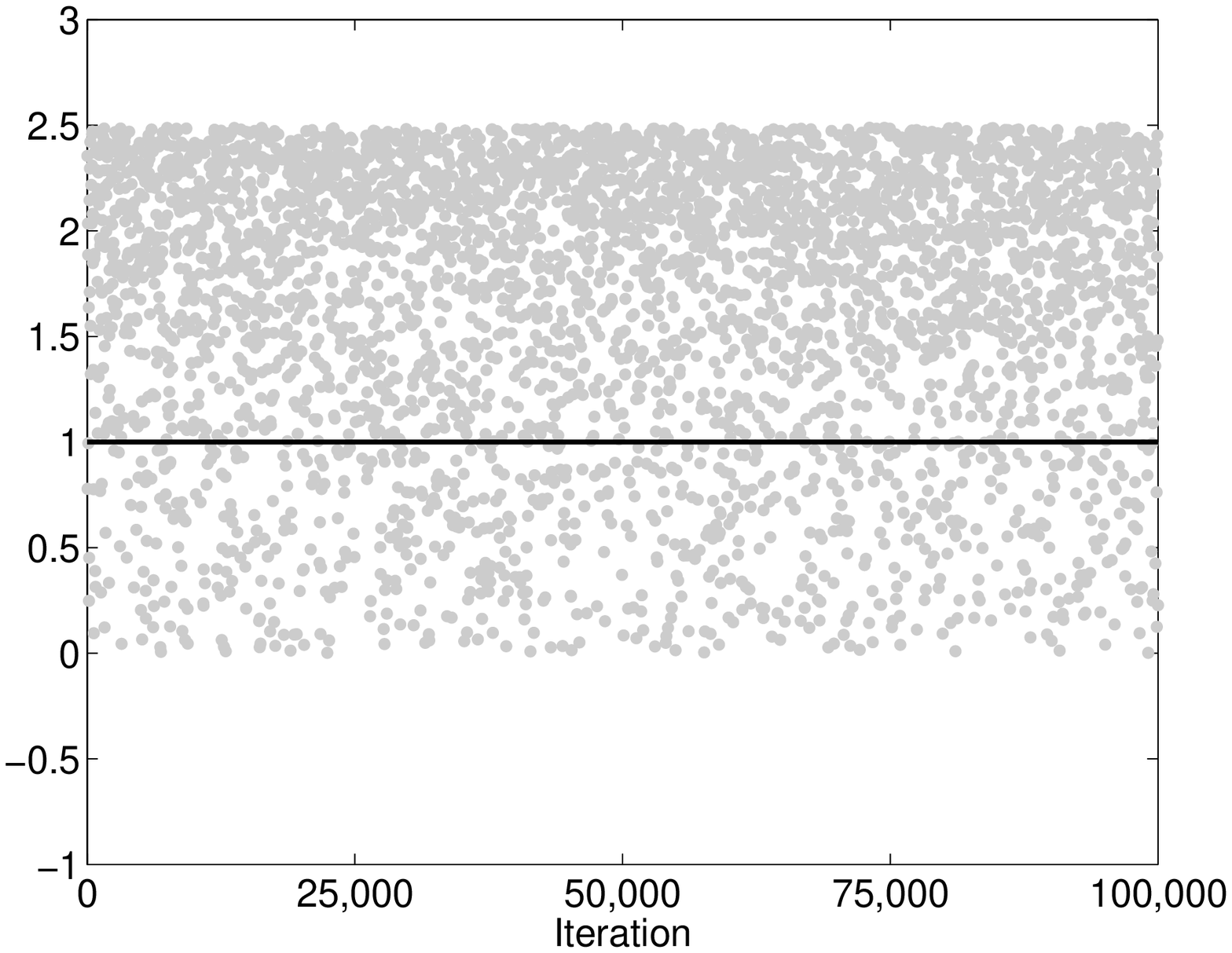}
                 \caption{$\eta_{1}$, Basic + all.}
		 \vspace{12pt}
         \end{subfigure}

         \centering
         \begin{subfigure}[b]{0.45\textwidth}
                 \includegraphics[width=\textwidth]{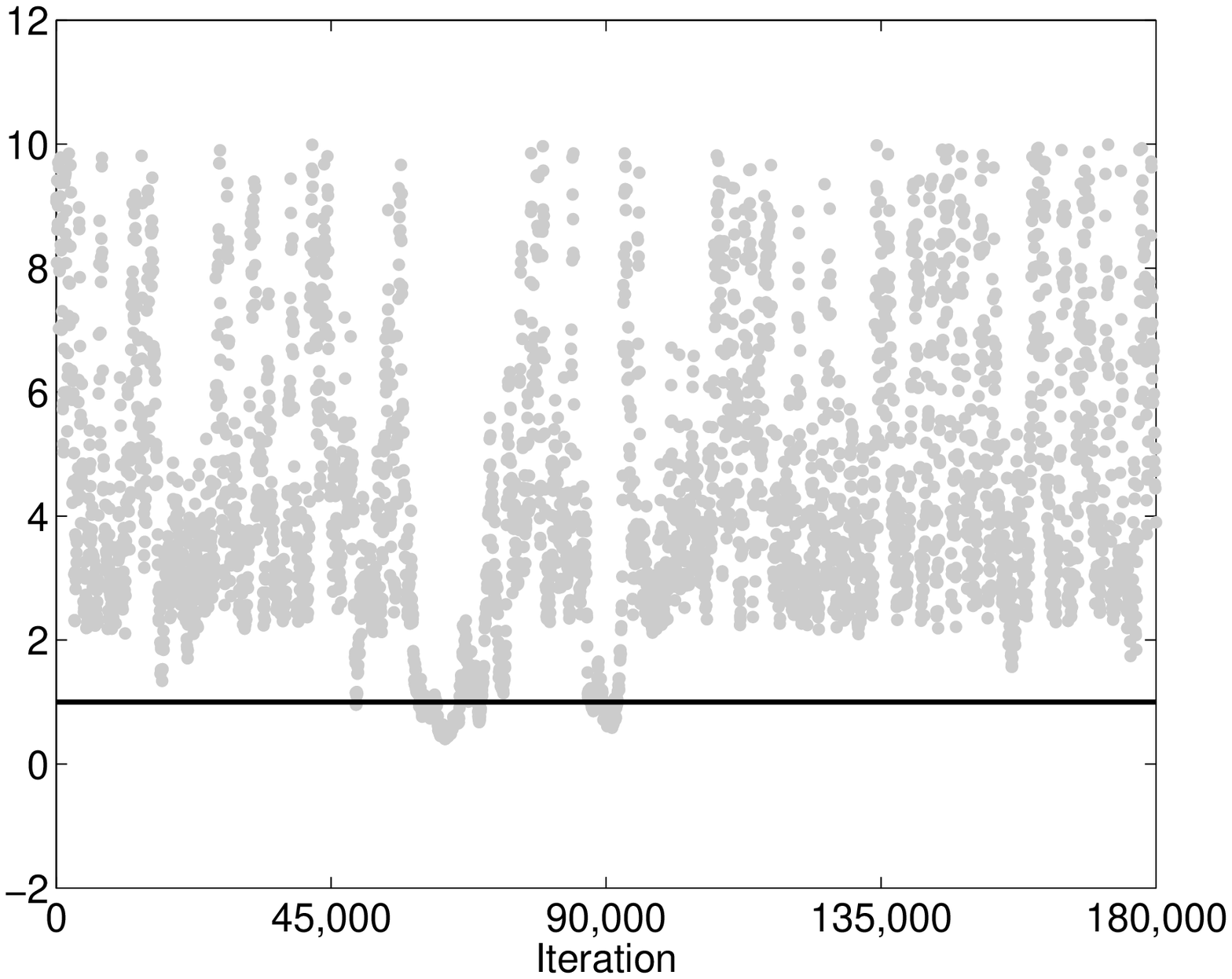}
                 \caption{$\eta_{2}$, Basic scheme.}
		 \vspace{12pt}
         \end{subfigure}
	~
         \begin{subfigure}[b]{0.45\textwidth}
                 \includegraphics[width=\textwidth]{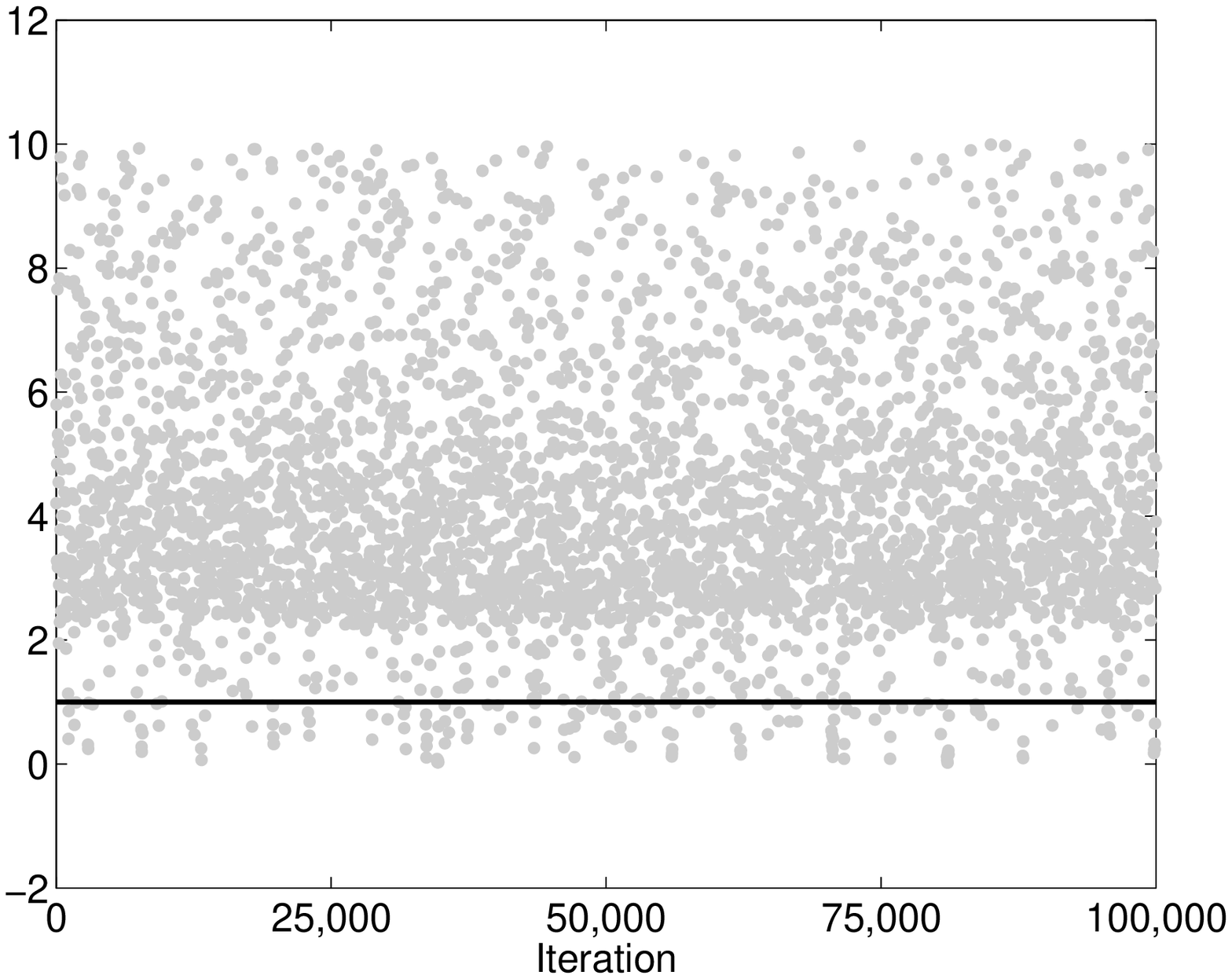}
                 \caption{$\eta_{2}$, Basic + all.}
		 \vspace{12pt}
         \end{subfigure}

         \centering
         \begin{subfigure}[b]{0.45\textwidth}
                 \includegraphics[width=\textwidth]{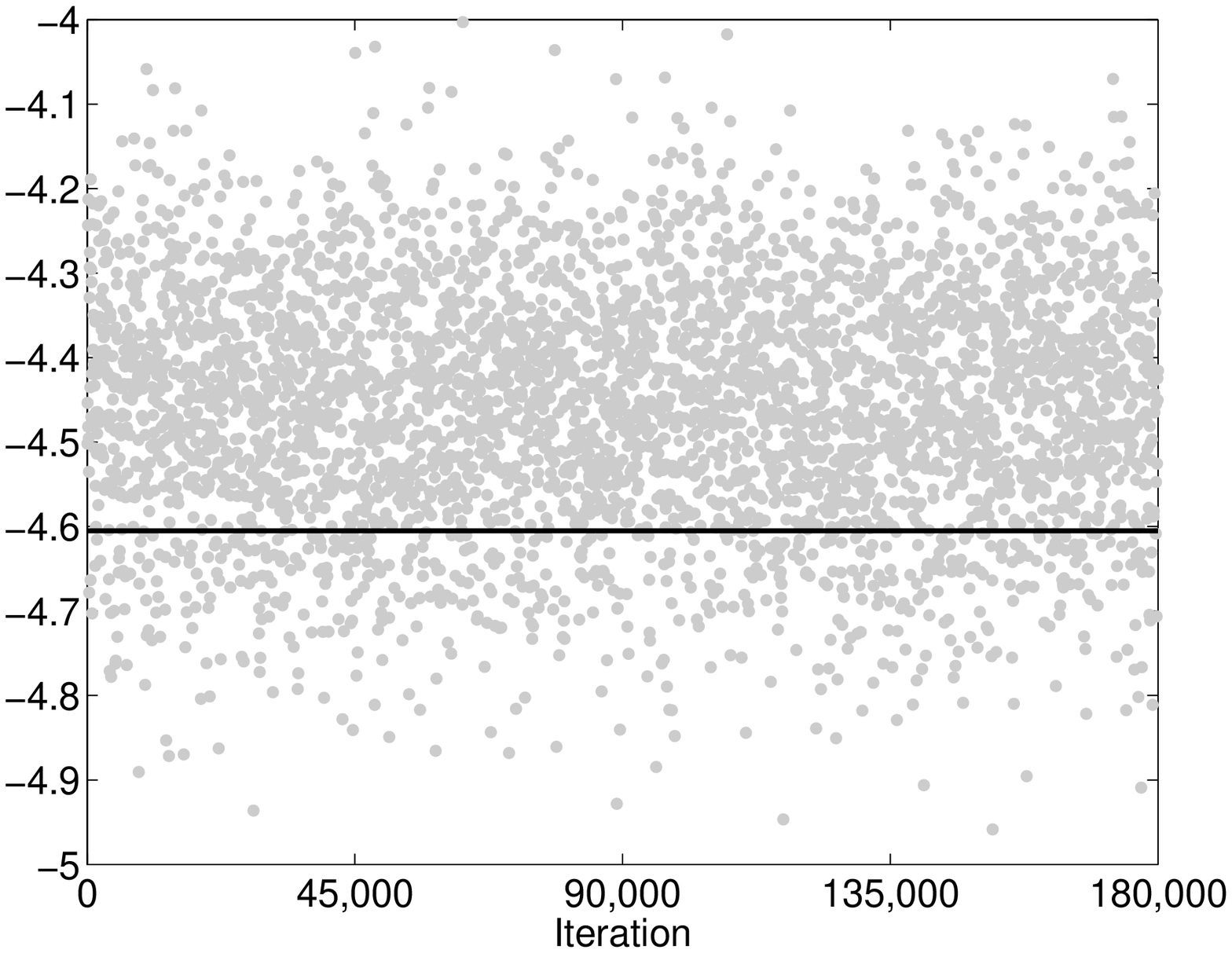}
                 \caption{$\eta_{3}$, Basic scheme.}
		 \vspace{12pt}
         \end{subfigure}
	~
         \begin{subfigure}[b]{0.45\textwidth}
                 \includegraphics[width=\textwidth]{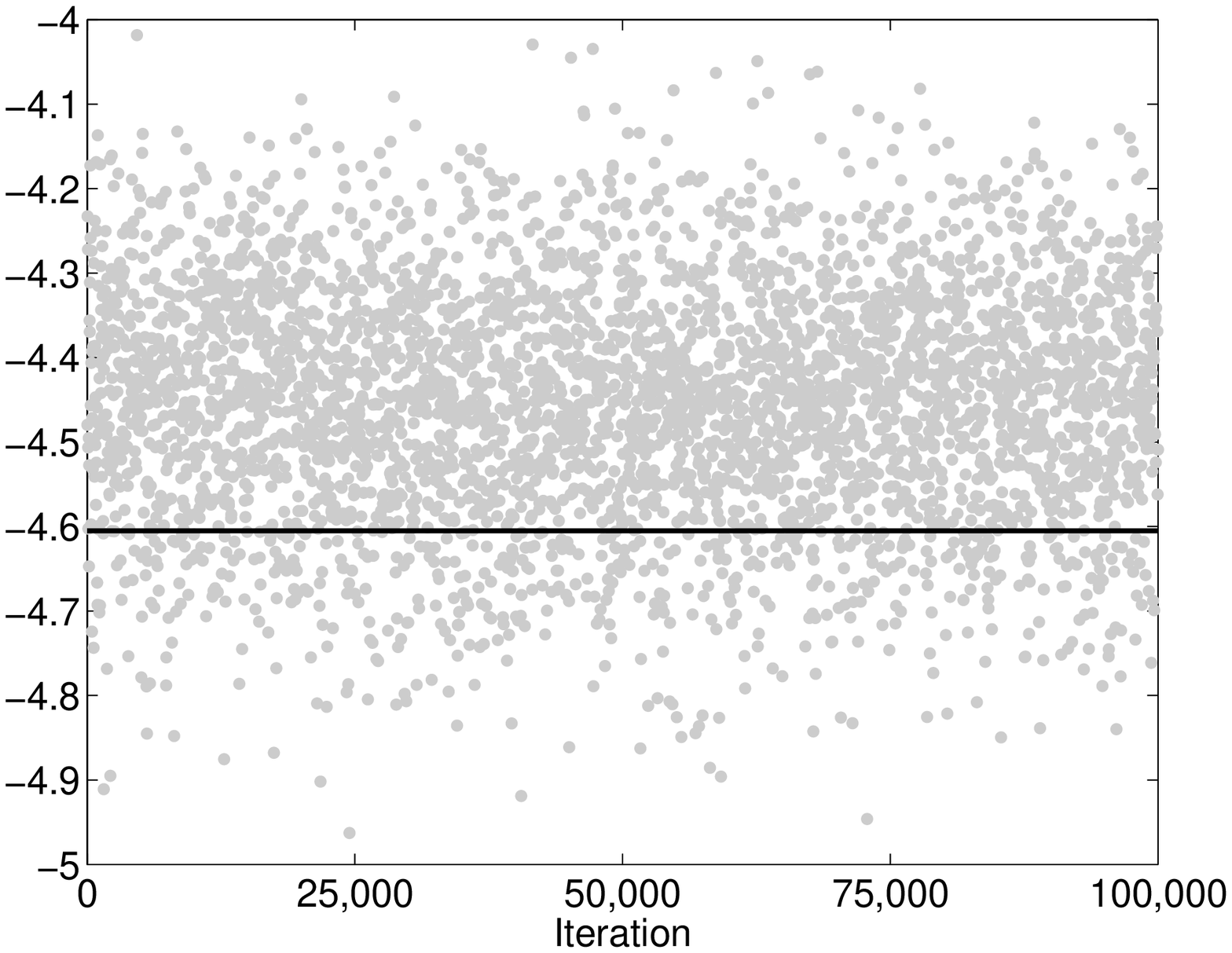}
                 \caption{$\eta_{3}$, Basic + all.}
		 \vspace{12pt}
         \end{subfigure}
         \caption{Comparison of performance for rare arrivals.}\label{fig:tracerare}

\end{figure}

We illustrate the performance of the samplers with trace plots in Figures \ref{fig:tracefreq}, \ref{fig:traceinter}, and \ref{fig:tracerare}. To produce these figures, we ran the samplers for an equal amount of computation time and thinned each run to $4,000$ points. The black line on each plot is the true parameter value.

In the case of frequent arrivals, the marginal posterior of $\eta_{3}$ is diffuse but concentrated given all $v_{i}$ and the data, so simultaneously updating all $v_{i}$ and $\eta_{3}$ makes sampling more efficient. In the case of rare arrivals, the marginal posteriors of both $\eta_{1}$ and $\eta_{2}$ are diffuse, while they are concentrated given all $v_{i}$ and the data. Simultaneously updating all $v_{i}$ and either $\eta_{1}$ or $\eta_{2}$ then leads to a noticeable gain in efficiency. In the other cases, the data is more informative about the parameters, so additional knowledge of the latent variables does not change the concentration of the posterior by as much. As a result, sampling efficiency is not affected as significantly by changing the latent variables and the parameters simultaneously.  

\section{Conclusion}

In this paper, we have shown how Bayesian inference with MCMC can be performed for the M/G/1 queueing model. As mentioned earlier, Fearnhead and Prangle (2010) used ABC for inference in the M/G/1 queueing model. Fearnhead and Prangle (2010) also used ABC for inference in the Ricker model of population dynamics, in which we assume that a population process is observed with Poisson noise. 

The basis of all ABC methods is the ability to simulate observations from a given stochastic model. The simulation process for a set of observations is always driven by a latent process of sampling and transforming certain random variables. The distributions of these random variables then determine the distribution of the final observed quantities. If we consider the latent variables which drive the simulation process jointly with the observations, then we can think of the observations as coming from a model with some latent structrure. These latent variables and the observed data will sometimes have a tractable joint density, which makes doing Bayesian inference with MCMC possible, at least in principle.

In an earlier work, Shestopaloff and Neal (2013), we used the same approach as in this paper to do Bayesian inference for the Ricker model, i.e. including additional latent variables in the MCMC state and sampling for them as well as model parameters.  We compared a basic MCMC scheme with several ``ensemble MCMC'' schemes for Bayesian inference in the Ricker model, and showed that using the ensemble schemes leads to a significant improvement in efficiency when compared to the basic MCMC scheme. Like for the M/G/1 queueing model, we have shown that Bayesian inference with MCMC is possible for the Ricker model, but requires more sophisticated MCMC methods for sampling to be efficient. 

It would be interesting to apply alternative inference methods for models with a time series structure to the M/G/1 queue, for example Particle Filters and Particle Markov Chain Monte Carlo (PMCMC) (Andrieu, Doucet, and Holenstein (2010)). 

As mentioned earlier, it should be possible to extend the MCMC method in this paper to service time distributions other than the Uniform one used in this paper. The most direct extension would be to consider location-scale service time distributions. Besides the simple Metropolis updates, we can then do additional updates for the location parameter (or some 1-to-1 function of it) using a shift update and additional updates for the scale parameter (or some 1-to-1 function of it) using a range scale update. Computation would not be impacted so long as (\ref{eq:logpost}) can be written in terms of low-dimensional sufficient statistics.

\section*{Acknowledgements}

This research was supported by the Natural Sciences and Engineering Research Council of Canada.  A.~S.\ is in part funded by an NSERC Postgraduate Scholarship. R.~N.\ holds a Canada Research Chair in Statistics and Machine Learning.

\section*{References}

\leftmargini 0.2in
\labelsep 0in

\begin{description}

\item
Andrieu, C., Doucet, A. and Holenstein, R. (2010), ``Particle Markov chain Monte Carlo methods''. {\em Journal of the Royal Statistical Society B}, vol.~72, pp.~269-342.

\item
Blum, M.G.B. and Francois, O. (2010). ``Non-linear regression models for Approximate Bayesian Computation'', {\em Statistics and Computing}, vol.~20, pp.~63-73.

\item
Bonassi, F.V. (2013) ``Approximate Bayesian Computation for Complex Dynamic Systems''. Ph.D. Thesis, Department of Statistical Science, Duke University.

\item
Fearnhead, P., Prangle, D. (2012) ``Constructing summary statistics for approximate Bayesian computation: semi-automatic approximate Bayesian computation'', {\em Journal of the Royal Statistical Society B}, vol.~74, pp.~1-28.

\item
Geyer, C. J. (2003). ``The Metropolis-Hastings-Green Algorithm'', \\ http://www.stat.umn.edu/geyer/f05/8931/bmhg.pdf

\item 
Metropolis, N., Rosenbluth, A.W., Rosenbluth, M.N., Teller, A.H., and Teller, E. (1953). ``Equation of State Calculations by Fast Computing Machines''. {\em Journal of Chemical Physics}, vol.~21, pp.~1087-1092.

\item
Neal, R. M. (1993) ``Probabilistic Inference Using Markov Chain Monte Carlo Methods'', Technical Report CRG-TR-93-1, Dept. of Computer Science, University of Toronto.

\item
Neal, R. M., Beal, M. J., and Roweis, S. T. (2004) ``Inferring state sequences for non-linear systems with embedded hidden Markov models'', in S. Thrun, et al (editors), {\em Advances in Neural Information Processing Systems 16}, MIT Press.

\item
Neal, R. M. (2003) ``Markov Chain Sampling for Non-linear State Space Models using Embedded Hidden Markov Models'', Technical Report No. 0304, Department of Statistics, University of Toronto, http://arxiv.org/abs/math/0305039.

\item
Shestopaloff, A. Y. and Neal, R. M. (2013). ``MCMC for non-linear state space models using ensembles of latent sequences'', Technical Report, http://arxiv.org/abs/1305.0320.

\end{description}

\end{document}